\pdfoutput=1

\PassOptionsToPackage{x11names,svgnames}{xcolor}

\documentclass[submission, Phys]{SciPost}
\hypersetup{
    colorlinks,
    linkcolor={red!50!black},
    citecolor={blue!50!black},
    urlcolor={blue!80!black}
}

\usepackage[bitstream-charter]{mathdesign}
\DeclareSymbolFont{usualmathcal}{OMS}{cmsy}{m}{n}
\DeclareSymbolFontAlphabet{\mathcal}{usualmathcal}

\fancypagestyle{SPstyle}{
\fancyhf{}
\lhead{\colorbox{scipostblue}{\bf \color{white} ~SciPost Physics}}
\rhead{{\bf \color{scipostdeepblue} ~Submission }}

\fancyfoot[C]{\textbf{\thepage}}
}

\usepackage{doi}
\usepackage{xspace}
\usepackage{graphicx}
\graphicspath{{Figures/}}
\usepackage{afterpage}
\usepackage{lineno}
\usepackage{tikz}
\usepackage{booktabs}
\usetikzlibrary{arrows,shapes,backgrounds,fit,positioning,calc}
\definecolor{irn-bru}{RGB}{239,129,0}

\usepackage{subfig}
\usepackage{caption}
\usepackage[section]{placeins}

\usepackage{microtype}
\makeatletter
\g@addto@macro\bfseries{\boldmath}
\makeatother

\usepackage{xpatch}
\usepackage{xparse}

\usepackage{listings}
\usepackage{siunitx}
\DeclareSIUnit{\electronvolt}{\text{e\kern-0.15ex V}}
\DeclareSIUnit{\eV}{\electronvolt}
\DeclareSIUnit{\MeV}{\mega\eV}
\DeclareSIUnit{\GeV}{\giga\eV}
\DeclareSIUnit{\TeV}{\tera\kern-0.1ex\eV}
\DeclareSIUnit{\ifb}{\femto\barn\tothe{-1}}

\usepackage{relsize}
\newcommand{\kbd}[1]{\text{\smaller \texttt{#1}}\xspace}

\NewDocumentCommand{\rivet}{o}{\text{R\protect\scalebox{0.8}{IVET}\IfNoValueTF{#1}{}{\,#1}}\xspace}
\NewDocumentCommand{\contur}{o}{\text{C\protect\scalebox{0.8}{ONTUR}\IfNoValueTF{#1}{}{\,#1}}\xspace}
\NewDocumentCommand{\yoda}{o}{\text{Y\protect\scalebox{0.8}{ODA}\IfNoValueTF{#1}{}{\,#1}}\xspace}
\newcommand{\Rivet}{\rivet\xspace}

\newcommand{\Professor}{P\protect\scalebox{0.8}{ROFESSOR}\xspace}

\newcommand{\Apfelpp}{A\protect\scalebox{0.8}{PFEL++}\xspace}
\newcommand{\Apfelppp}{\Apfelpp\ensuremath{^\prime}\xspace}

\newcommand{\Minuit}{M\protect\scalebox{0.8}{INUIT}\xspace}

\newcommand{\muF}{\ensuremath{\mu_\mathrm{F}}\xspace}
\newcommand{\muR}{\ensuremath{\mu_\mathrm{R}}\xspace}

\newcommand{\MCatNLO}{M\protect\scalebox{0.8}{C}@N\protect\scalebox{0.8}{LO}\xspace}

\newcommand{\Collier}{C\protect\scalebox{0.8}{OLLIER}\xspace}

\newcommand{\OpenLoops}{O\protect\scalebox{0.8}{PEN}L\protect\scalebox{0.8}{OOPS}\xspace}

\newcommand{\Sherpa}{S\protect\scalebox{0.8}{HERPA}\xspace}
\newcommand{\Comix}{C\protect\scalebox{0.8}{OMIX}\xspace}

\newcommand{\Amegic}{A\protect\scalebox{0.8}{MEGIC}\xspace}
\newcommand{\CSS}{CSS\protect\scalebox{0.8}{HOWER}\xspace}

\newcommand{\Ahadic}{A\protect\scalebox{0.8}{HADIC}\xspace}

\newcommand{\Amisic}{A\protect\scalebox{0.8}{MISIC}\xspace}
\newcommand{\Pythia}{P\protect\scalebox{0.8}{YTHIA}\xspace}

\newcommand{\LHAPDF}{L\protect\scalebox{0.8}{HAPDF}\xspace}

\newcommand{\pdfforlhc}{\ensuremath{\text{PDF4LHC21}}\xspace}

\newcommand{\plgamma}{\ensuremath{\text{PL}\gamma26}\xspace}

\newcommand\lep{L\scalebox{0.85}{EP}\xspace}

\newcommand\LEP{\lep}

\newcommand\HERA{H\scalebox{0.8}{ERA}\xspace}

\newcommand{\done}{{\mathrm d}}

\newcommand{\mc}[1]{\mathcal{#1}}
\newcommand{\mb}[1]{\mathbf{#1}}

\newcommand{\bea}{\begin{align}}
\newcommand{\eea}{\end{align}}
\newcommand{\beq}{\begin{equation}}
\newcommand{\eeq}{\end{equation}}

\newcommand{\bs}{\begin{split}}
\newcommand{\es}{\end{split}}

\newcommand{\bi}{\begin{itemize}}
\newcommand{\ei}{\end{itemize}}
\newcommand{\bc}{\begin{center}}
\newcommand{\ec}{\end{center}}
\newcommand{\bac}{\begin{array}{c}}
\newcommand{\bacc}{\begin{array}{cc}}
\newcommand{\ea}{\end{array}}

\def\spa#1.#2{\langle#1\,#2\rangle}
\def\spb#1.#2{[#1\,#2]}

\newcommand{\MSbar}{\ensuremath{\overline{\mathrm{MS}}}\xspace}
\newcommand{\DISgamma}{\ensuremath{{\mathrm{DIS}}_\gamma}\xspace}

\newcommand{\alphaS}{\ensuremath{\alpha_\text{s}}\xspace}
\newcommand{\alphaem}{\ensuremath{\alpha_\text{em}}\xspace}
\newcommand{\alphaSn}[1]{\ensuremath{\alpha^#1_\text{s}}\xspace}

\newcommand{\pvec}{\ensuremath{\vec{p}}\xspace}
\newcommand{\pvectune}{\ensuremath{\vec{p}_\mathrm{tune}}\xspace}

\newcommand{\qzero}{\ensuremath{Q_0}\xspace}

\newcommand{\xgamma}{\ensuremath{x_\gamma}\xspace}
\newcommand{\xgammaobs}{\ensuremath{x_\gamma^{\mathrm{obs}}}\xspace}

\newcommand{\qsquared}{\ensuremath{Q^2}\xspace}
\newcommand{\pdf}{\ensuremath{f_i^{(p)}(x,\qsquared)}\xspace}
\newcommand{\photonpdf}{\ensuremath{f_i^{(\gamma)}(x,\qsquared)}\xspace}
\newcommand{\kgamma}{\ensuremath{\mb{k}_\gamma}}
\newcommand{\pij}{\ensuremath{\mb{P}_{ij}}\xspace}
\newcommand{\cgamma}{\ensuremath{C_\gamma}\xspace}
\newcommand{\ftwogamma}{\ensuremath{F_2^\gamma}\xspace}

\newcommand{\apl}{\ensuremath{A_{\mathrm{PL}}}\xspace}
\newcommand{\bpl}{\ensuremath{B_{\mathrm{PL}}}\xspace}
\newcommand{\ahad}{\ensuremath{A^{\mathrm{had}}}\xspace}
\newcommand{\bhad}{\ensuremath{B^{\mathrm{had}}}\xspace}
\newcommand{\chad}{\ensuremath{C^{\mathrm{had}}}\xspace}
\newcommand{\ahadg}{\ensuremath{A^{\mathrm{had}}_G}\xspace}
\newcommand{\bhadg}{\ensuremath{B^{\mathrm{had}}_G}\xspace}
\newcommand{\chadg}{\ensuremath{C^{\mathrm{had}}_G}\xspace}

\newcommand{\sigmaNDnorm}{\ensuremath{\sigma^{\mathrm{ND}}_{\mathrm{norm}}}\xspace}
\newcommand{\pTnought}{\ensuremath{p_{\mathrm{T},0}}\xspace}
\newcommand{\pTnoughtref}{\ensuremath{p_{\mathrm{T},0}^{\mathrm{ref}}}\xspace}
\newcommand{\pTmin}{\ensuremath{p_{\mathrm{T},\mathrm{min}}}\xspace}
\newcommand{\pTminref}{\ensuremath{p_{\mathrm{T},\mathrm{min}}^{\mathrm{ref}}}\xspace}
\newcommand{\pTnoughtIR}{\ensuremath{p_{\mathrm{T},0}^{\mathrm{IR}}}\xspace}
\newcommand{\pTminIR}{\ensuremath{p_{\mathrm{T},\mathrm{min}}^{\mathrm{IR}}}\xspace}
\newcommand{\Ecms}{\ensuremath{E_{\mathrm{CMS}}}\xspace}
\newcommand{\Ecmsref}{\ensuremath{E_{\mathrm{CMS}}^{\mathrm{ref}}}\xspace}

\makeatletter
\let\@oldto\to
\renewcommand{\to}{\ensuremath{\@oldto}\xspace}
\renewcommand{\to}{\ensuremath{\@oldto}\xspace}

\makeatother

\renewcommand{\vec}[1]{\mathbf{#1}}
\renewcommand{\d}{\mathrm{d}}

\NewDocumentCommand{\varcoord}{o}{\ensuremath{\IfNoValueTF{#1}{\vec{\omega}}{\omega^{(#1)}}}\xspace}

\NewDocumentCommand{\bincoord}{o}{\ensuremath{\IfNoValueTF{#1}{\vec{\theta}}{\theta^{(#1)}}}\xspace}

\NewDocumentCommand{\unbincoord}{o}{\ensuremath{\IfNoValueTF{#1}{y}{y}}\xspace} %

\newcommand{\pT}{\ensuremath{p_\mathrm{T}}\xspace}
\newcommand{\ET}{\ensuremath{E_\mathrm{T}}\xspace}

\hypersetup{hidelinks,
  pdfauthor={Andy Buckley, Frank Krauss, Peter Meinzinger, Vithyaban Anjelo Narendran},
  pdftitle={Particle-level parton densities and uncertainties for the resolved photon at NLO}
}

\begin{document}
\pagestyle{SPstyle}

\begin{flushright}
  \small
  IPPP/26/77\\
  MCnet-26-34\\
  ZU-TH 43/26
\end{flushright}

\begin{center}
  \color{scipostdeepblue}
  \Large
  \textbf{Particle-level parton densities and uncertainties\\ for the resolved photon at NLO}
\end{center}

\begin{center}
  Andy~Buckley$^1$,
  Frank~Krauss$^2$,
  Peter~Meinzinger$^3$,
  Vithyaban~Anjelo~Narendran$^{4*}$
\end{center}

\begin{center}
  \itshape
  $^1$ School of Physics \& Astronomy, University of Glasgow,\\ University~Place, G12~8QQ, Glasgow, UK\\
  $^2$ Institute for Particle Physics Phenomenology, Durham University\\ South Road, Durham, DH1~3LE, UK\\
  $^3$ Physik-Institut, Universit{\"a}t Z{\"u}rich, CH-8057 Z{\"u}rich, Switzerland\\
  $^4$ Department of Physics, University of Oxford, Denys Wilkinson Building, \\ Keble Road, Oxford, OX1~3RH, UK\\
\end{center}

\begin{center}
\today
\end{center}

\section*{Abstract}
We present a new set of parton distribution functions (PDFs), \plgamma, for the resolved photon at next-to-leading order (NLO), the first to be fitted to a wide range of jet measurements from several experiments.
Jet production in photoproduction at \HERA and in $\gamma\gamma$ collisions at \LEP is sensitive to both the quark and the gluon content of the photon, but the measurements lie at low scales, where non-perturbative corrections are large.
We therefore choose a novel approach and fit the PDFs to these measurements at particle level, using NLO predictions from the general-purpose event generator \Sherpa, matched to the parton shower and including hadronisation and multi-parton interactions (MPI).
As part of the procedure, we tune the MPI model separately for photon--proton and photon--photon collisions.
Modelling these non-factorisable effects in the fit, rather than through external correction factors, lets us include many more measurements and data points.
We estimate the experimental uncertainties with replicas and compare the fitted PDFs with previous sets.

\section{Introduction}
\label{sec:intro}

Resolved, i.e.~hadronic, photons are a wonderfully counterintuitive
consequence of QCD. Despite being devoid of colour charge, the
electromagnetic coupling of photons to quarks enables seeding of a
recursive QCD evolution from which hadronic structure emerges in these
very unhadronic particles.

The study of resolved-photon structure, which had produced a series of
LO and NLO photon-PDF
sets~\cite{Gluck:1991jc,Schuler:1995fk,Cornet:2004nb,Aurenche:2005da,Slominski:2005bw},
essentially paused with the end of \HERA operations in 2007, and in the intervening 20 years the focus
of the high-energy physics PDF community has instead been on proton
PDFs for (and constrained by) LHC data analyses. However, interest is
resurgent in anticipation of the Electron--Ion Collider (EIC),
scheduled to begin data-taking at Brookhaven in the
2030s~\cite{AbdulKhalek:2021gbh}.
The recent VALO\,1.0 fits~\cite{Chithirasreemadam:2026mqp} are an example of this renewed
interest: they use historical $F_2^\gamma$ measurements from
$e^+e^-$ colliders~\cite{Nisius:1999cv} to produce the first new published photon-PDF sets since
2006~\cite{Slominski:2005bw}. Fits to $F_2^\gamma$ are the theoretically cleanest and
best-established route to the photon's parton content, since the
tagged lepton fixes the probed scale~\cite{Krawczyk:2000mf,Gehrmann:2025wbc}.

In this paper we revisit \HERA and \LEP measurements of jet production
in photoproduction and $\gamma\gamma$ collisions.
These measurements are numerous and, unlike $F_2^\gamma$, probe
the gluon content of the photon directly.
As these measurements were typically done at relatively low scales,
they are however also subject to large non-perturbative corrections.

Two developments of the past two decades make it worthwhile to revisit
these data. First, the much greater precision and technical capability
of modern particle-level event generators make it tractable to fit
PDFs not to fixed-order partonic calculations but directly to
particle-level final states via a shower+hadronisation event
generator, so that non-perturbative effects enter the fit itself
rather than as external correction factors. Second, PDF error sets
have become the standard means of quantifying PDF uncertainties, and
the fit presented here provides the first one for the resolved photon from jet data. Using
on-the-fly PDF reweighting and a \Professor-style surrogate model~\cite{Buckley:2009bj}, the
particle-level fit comes at modest computational cost. As the
multi-parton-interaction (MPI) model of the generator is untuned for
photon-induced collisions, and MPI effects could otherwise be absorbed
into the fitted PDFs, we first tune it to the same data; the resulting
MPI tunes for \HERA and \LEP are a further outcome of this study.
We describe the technical process and results of these
particle-level PDF (PL-PDF) fits, and discuss future prospects for
this approach.

The paper is organised as follows: Section~\ref{sec:plfit} introduces
the particle-level fitting strategy, Section~\ref{sec:processes} the
photoproduction and $\gamma\gamma$ processes and the separation of their
direct and resolved components, and Section~\ref{sec:theory} the
photon-PDF evolution and parametrisation. Section~\ref{sec:evgen}
describes the event generation, including the MPI model, and
Section~\ref{sec:method} the fitting methodology. We present the results
in Section~\ref{sec:results}, discuss prospects in Section~\ref{sec:further},
and conclude in Section~\ref{sec:concl}.

\section{Particle-level PDF fits}
\label{sec:plfit}

PDFs are indirectly defined quantities, being not standalone
observables but functions related to observables by convolution with a
scattering-amplitude function over permitted initial-state
configurations. The classic presentation of this is via the
factorisation theorem~\cite{Collins:1989gx},
\begin{equation}
  \label{eq:factorisation}
  \done{\sigma({A B \to n})}
  =
  \sum_{a,b}
  \, \int_{0}^{1} \! \done{x_a}
  \, \int_{0}^{1} \! \done{x_b}
  ~
  f_a^{(A)}(x_a,\muF^2) \; f_b^{(B)}(x_b,\muF^2) \;
  \done\hat\sigma_{ab\to n}(\Phi, \muR, \muF) \, ,
\end{equation}
in which all parton flavours $a,b$ are extracted from initial-state
hadrons $A,B$, with all possible momentum fractions $x_a, x_b$
respectively, provided they are compatible with the set of
hard-scattering amplitude functions $\done\hat\sigma$ mapping to
final-state $n$.

In principle, the final state $n$ could be fully exclusive
event-topologies as observed at colliders, if there was a full and tractable
QFT model through to hadron-level states; in practice, the
current limitation is to a partonic final state of very limited
multiplicity. The consequence is that fits of PDFs defined under this
model, specific to a partonic matrix-element function at fixed
loop-order, must be fitted to collider data via theory predictions
that are ``corrected'' from fixed-order parton level to particle level
according to some model of radiative corrections and non-perturbative
hadronisation, hadron-decay, and multiple-interaction effects. In
practice, these correction factors are obtained by use of a
particle-level MC generator.

This suggests an alternative picture: rather than restrict to data for
which such a non-perturbative correction has been estimated, fit
instead to particle-level observables using modern MC-generator tools
in which the matrix-element is augmented with radiative corrections
via QCD parton-shower algorithms, and generative hadronisation,
multiple-interaction and decay models. The loop-order of matrix element
supported by this scheme is lower than for the fixed-order approach
--- currently a maximum of 1- or sometimes 2-loop shower-matched
processes as opposed to 2- to 3-loop for traditional PDF fits --- but
this can be traded against the ability to use a wider range of
experimental data, and the fact that a major use of fitted PDFs is as
part of the lower-order MC-generation chain. For the resolved photon,
the trade-off is particularly favourable: the historical photon-PDF
fits, up to and including SAL~\cite{Slominski:2005bw}, are at most NLO, which
is also the accuracy of the \MCatNLO-matched predictions used here, cf.\
Section~\ref{sec:evgen}. As differential
cross-sections can depend on both the loop- and leg-orders of the
scattering process, which typically are respectively lower and higher
for ME+PS generators than fixed-order, the current paradigm results in
PDF functions that are not obviously optimal for MC-generator
predictivity. Dedicated PDF sets for LO event generators, such as MRST
LO$^*$~\cite{Sherstnev:2007nd,Sherstnev:2008dm} and CT09MC~\cite{Lai:2009ne}, address this
in part, but are still fitted to fixed-order theory. The PDFs fitted
here use a different approach: they are genuine NLO PDFs in the \MSbar
scheme with standard DGLAP evolution and can be used in fixed-order
calculations. What differs from traditional fits is how non-factorisable
corrections are included: instead of ad-hoc external corrections applied
to the raw data, we include them on-the-fly using tuned models.

\subsection{Logistics of particle-level PDF fits}

To our knowledge, particle-level PDF fits have not previously been
performed, with the practical obstacle being computational tractability. The na\"ive approach to
such a fit would be to run a fit-optimisation algorithm in the space
of the PDF parametrisation, with each proposed PDF being used to
generate millions of particle-level events and hence compute
statistically stable expected observables, before making the next PDF
proposal based on their comparison to experimental data. Even with
parallel ensembles of optimisers, this approach is intractable due to
the multiple hours or days of generation time needed in each serial
step. Nor can the interpolated coefficient grids that make fixed-order
PDF fits fast~\cite{Kluge:2006xs,Britzger:2012bs,Carli:2010rw} be used directly, as the modelling goes
beyond fixed order.

However, this limitation has long been worked
around for the ``tuning'' optimisation of generators' own
physics-model parameters, in particular via the \Professor toolkit's
approach~\cite{Buckley:2009bj}. In this framework, an ensemble of efficient algebraic surrogates, or
emulators, models each observable-bin value as a function of the
model parameters, with established applications at
\LEP~\cite{DELPHI:1996sen} and the LHC~\cite{TheATLAScollaboration:2014rfk}.
Replacing the generator-parameters of the \Professor methodology with
the parameters of a closed-form PDF set requires only minor
modification to the established MC-tuning methodology, allowing an
MC-coupled PDF set to be ``tuned''.

MC tuning typically requires a large up-front (but trivially
parallelisable) MC-generation campaign to train the surrogates, as the
focus of tuning is on phenomenological components of the models which
require explicit re-runs of the generator for each parameter
point.\footnote{Though this situation is changing with the increasing
implementation of on-the-fly event-weighting technology for
hadron-level physics~\cite{Pabst:2026pwv,Bierlich:2023fmh}.} For PDF fits, however, a
more performant approach is available, as general-purpose
event generators support propagation of multiple event-weight streams for
PDF variations.\footnote{At least in the matrix-element sampling and
evaluation, accounting for PDF differences in NLO counter-terms; some
generators, including \Sherpa, also propagate PDF variations for
additional emissions via the initial-state parton shower.} Subject to
the usual concerns of statistical support for event-reweighting, the
\emph{a priori} PDF proposals for surrogate-building can be created
via single MC-generator runs with a large set of PDF weights, expanded
into an ensemble of MC-observable datasets via the \Rivet
package~\cite{Bierlich:2024vqo}.

\section{Resolved-photon processes}\label{sec:processes}

The collinear emission of a quasi-real photon can be computed in the Equivalent Photon Approximation (EPA), for electrons also known as the Weizsäcker--Williams formula~\cite{vonWeizsacker:1934nji,Williams:1934ad,Budnev:1975poe}. It is based on the observation that photons with small virtualities, $Q^2 < Q^2_\mathrm{max}$, can be approximated by real photons, where the virtuality can be integrated over.
In terms of the total cross-section this equates to factorising the electron-photon vertex, necessitating a convolution with this flux, additional to the regular PDFs.
We use an improved version of the spectrum, following~\cite{Frixione:1993yw}, which uses a correction for large $x$, resulting in the following expression:
\begin{equation}
    f^{(e)}_\gamma (x) = \frac{\alpha_\mathrm{em}}{2 \pi x}
                \left[ \left( 1 + (1 - x)^2 \right) \log \left( \frac{Q^2_\mathrm{max}}{Q^2_\mathrm{min}} \right) -
                2 m_e^2 x^2 \left( \frac{1}{Q^2_\mathrm{min}} - \frac{1}{Q^2_\mathrm{max}} \right) \right] \label{eq:photonflux} \, .
\end{equation}
Here, $x$ denotes the ratio of photon to electron energy, $E_\gamma/E_e$, $\alpha_\mathrm{em}$ is the electromagnetic coupling constant and
$Q^2_\mathrm{max}$ and $Q^2_\mathrm{min}$ denote the maximal and minimal photon virtuality. The former is effectively a phase-space cut given by the measurement, and the latter can be calculated from kinematic restrictions as
\begin{equation}
    Q^2_\mathrm{min} = \frac{m_e^2 x^2}{1 - x} \ .
\end{equation}

\noindent
For jet production in photon scattering, we have to take into account both the direct and the resolved part to arrive at a physically meaningful cross-section.

In the \emph{direct} component, the quasi-real photon couples as a whole, i.e.\ as a point-like particle, to the hard interaction.
In the \emph{resolved} component, it first fluctuates into a hadronic state, out of which a parton $i$ carrying a fraction $x'$ of the photon momentum enters the hard interaction.
The latter is described by the photon PDFs $f^{(\gamma)}_i(x', \muF^2)$, which are the subject of this study.
The total cross-section is the sum of these components, each of which is written as a convolution of the photon flux(es), the relevant PDFs and the partonic cross-section~$\hat{\sigma}$.

\paragraph{Photoproduction at \HERA.}
In electron-proton collisions, a photon emitted by the electron scatters off a parton of the proton, and the cross-section splits into a direct and a resolved component,
\begin{equation}
    \sigma \left( e p \to e X \right) = \sigma_\text{dir} + \sigma_\text{res} \ ,
\end{equation}
with
\begin{align}
    \sigma_\text{dir} =& \sum_{j}
    \int_{0}^{1} \d x_1 \, f^{(e)}_\gamma \left( x_1 \right)
    \int_{0}^{1} \d x_2 \, f^{(p)}_j \left( x_2, \muF^2 \right) \,
    \hat{\sigma} \left( \gamma j \to X \right) \ , \\
    \sigma_\text{res} =& \sum_{i,j}
    \int_{0}^{1} \d x_1 \, f^{(e)}_\gamma \left( x_1 \right)
    \int_{0}^{1} \d x'_1 \, f^{(\gamma)}_i \left( x'_1, \muF^2 \right)
    \int_{0}^{1} \d x_2 \, f^{(p)}_j \left( x_2, \muF^2 \right) \,
    \hat{\sigma} \left( i j \to X \right) \ .
\end{align}
Here, $x_1$ is the momentum fraction of the photon with respect to the electron, which coincides with the inelasticity $y$ in Eq.~(\ref{eq:xgammaobs}), $x'_1$ is the momentum fraction of the parton with respect to the photon, and $x_2$ is the momentum fraction of the parton with respect to the proton, described by the proton PDFs $f^{(p)}_j$.

\paragraph{Photon-photon collisions at \LEP.}
In $e^+e^-$ collisions, both leptons emit a quasi-real photon, each of which can either interact directly or be resolved.
Correspondingly, the cross-section is the sum of a direct, two single-resolved, and a double-resolved component,
\begin{equation}
    \sigma \left( e e \to e e X \right) = \sigma_\text{dir} + \sigma_\text{SR,1} + \sigma_\text{SR,2} + \sigma_\text{DR} \ ,
\end{equation}
with
\begin{align}
    \sigma_\text{dir} =&
    \int_{0}^{1} \d x_1 \, f^{(e)}_\gamma \left( x_1 \right)
    \int_{0}^{1} \d x_2 \, f^{(e)}_\gamma \left( x_2 \right) \,
    \hat{\sigma} \left( \gamma \gamma \to X \right) \ , \\
    \sigma_\text{SR,1} =& \sum_{i}
    \int_{0}^{1} \d x_1 \, f^{(e)}_\gamma \left( x_1 \right)
    \int_{0}^{1} \d x_2 \, f^{(e)}_\gamma \left( x_2 \right)
    \int_{0}^{1} \d x'_1 \, f^{(\gamma)}_i \left( x'_1, \muF^2 \right) \,
    \hat{\sigma} \left( i \gamma \to X \right) \ , \\
    \sigma_\text{SR,2} =& \sum_{j}
    \int_{0}^{1} \d x_1 \, f^{(e)}_\gamma \left( x_1 \right)
    \int_{0}^{1} \d x_2 \, f^{(e)}_\gamma \left( x_2 \right)
    \int_{0}^{1} \d x'_2 \, f^{(\gamma)}_j \left( x'_2, \muF^2 \right) \,
    \hat{\sigma} \left( \gamma j \to X \right) \ , \\
    \sigma_\text{DR} =& \sum_{i,j}
    \int_{0}^{1} \d x_1 \, f^{(e)}_\gamma \left( x_1 \right)
    \int_{0}^{1} \d x_2 \, f^{(e)}_\gamma \left( x_2 \right) ~\cdot \nonumber \\
    & \qquad \int_{0}^{1} \d x'_1 \, f^{(\gamma)}_i \left( x'_1, \muF^2 \right)
    \int_{0}^{1} \d x'_2 \, f^{(\gamma)}_j \left( x'_2, \muF^2 \right) \,
    \hat{\sigma} \left( i j \to X \right) \ ,
\end{align}
where $x_{1,2}$ denote the momentum fractions of the photons with respect to the leptons and $x'_{1,2}$ those of the partons with respect to the photons.
While the direct component is independent of the photon PDFs, the single-resolved components are linear and the double-resolved component is quadratic in them.

Beyond leading order, this decomposition is not unique.
The direct component contains collinear $\gamma \to q\bar{q}$ splittings, whose singularities are absorbed into the point-like part of the photon PDFs at the factorisation scale, \muF.
Hence, the individual components depend on \muF and on the factorisation scheme, and only their sum is a physical observable, in which this dependence cancels up to higher orders.
Throughout, we use the \MSbar factorisation scheme, both in the evolution of the photon PDFs and in the NLO calculations.

\subsection{Separating the resolved and direct components}

The \HERA and \LEP measurements used in our fit distinguish the direct and resolved components through the fraction \xgamma of the photon momentum entering the hard interaction.
At leading order, $\xgamma = 1$ for direct photons and $\xgamma < 1$ for resolved ones.
As \xgamma is not directly measurable, the \HERA experiments reconstruct the proxy \xgammaobs from the transverse energies $E_{\mathrm{T},i}^{\mathrm{jet}}$ and pseudorapidities $\eta_i^{\mathrm{jet}}$ of the $n$ leading jets, the inelasticity $y$, and the electron beam-energy $E_e$, as
\begin{equation}
  x_\gamma^{\mathrm{obs}} = \frac{\sum_{i=1}^{n} E_{\mathrm{T},i}^{\mathrm{jet}} \; \exp\left(-\eta_i^{\mathrm{jet}}\right)}{2yE_e} \, .
  \label{eq:xgammaobs}
\end{equation}

In the H1~\cite{H1:2002apm,H1:2006rre} and ZEUS~\cite{ZEUS:1997fwn,ZEUS:2001zoq,ZEUS:2007njl} dijet analyses used here,
\xgammaobs is built from the two jets with the highest transverse energy, i.e.\ $n=2$ in Eq.~(\ref{eq:xgammaobs});
the three- and four-jet analysis \texttt{ZEUS\_2007\_I756660}~\cite{ZEUS:2007xwd} uses $n=3$ or $4$, depending on the observable.
The inclusive-jet and charged-particle analyses do not use \xgammaobs.
The \LEP dijet analysis uses the analogous variables
\begin{equation}
  x_\gamma^\pm = \frac{\sum_{\mathrm{jets}} (E \pm p_z)}{\sum_{\mathrm{hadrons}} (E \pm p_z)} \, ,
  \label{eq:xgammapm}
\end{equation}
where the numerator sums over the two leading jets and the denominator over the full hadronic final state.

Additional radiation outside the jets and hadronisation shift direct events to values below one, so that \xgammaobs equals \xgamma only at leading order.
A cut at $\xgammaobs \approx 0.75$--$0.8$ conventionally separates a direct-enriched region above from a resolved-enriched region below.
As the decomposition into direct and resolved components is itself not unique beyond leading order, these regions are enriched in, but never pure samples of, either component.
Unlike \xgamma, \xgammaobs is built from jets and is well defined at any order and at particle level.
Many measurements are presented in separate \xgammaobs regions.

However, \xgammaobs is known to receive large non-perturbative corrections~\cite{Helenius:2024rth}.
In fixed-order fits, these effects enter as external correction factors, typically derived from leading-order MC generators.
In our particle-level approach, they are part of the simulation itself, through the hadronisation and MPI models, cf.~Section~\ref{sec:evgen}.

\section{Photon-PDF evolution}
\label{sec:theory}

Parton distributions in the case of the well-known proton PDFs satisfy the
homogeneous evolution equation
\begin{equation}
  \frac{\done \pdf}{\done \ln{\qsquared}} = \sum_j \pij \ast f_j^{(p)}(x,\qsquared) \, ,
  \label{eq:proton_evol}
\end{equation}
where the $\ast$ denotes the convolution of the purely hadronic
splitting function \pij (including all coupling factors)
with the PDF $f_j^{(p)}$ in Bjorken-$x$ space:
\begin{equation}
  \pij \ast f_j^{(p)} = \int_x^1 \frac{\done z}{z} \, \pij(z, \qsquared) \, f_j^{(p)}\!\left(\frac{x}{z}, \qsquared\right) .
  \label{eq:convolution}
\end{equation}
The case of the photon is more interesting, as the $\gamma \rightarrow q\bar{q}$
splitting introduces an additional source term $k_i$, resulting in the inhomogeneous
equation
\begin{equation}
  \frac{\done \photonpdf}{\done \ln{\qsquared}} = \sum_j \pij \ast f_j^{(\gamma)}(x,\qsquared) + k_i(x,\qsquared) \, .
  \label{eq:photon_evol}
\end{equation}
We may retain the QCD splitting functions of \pij, and the photon-parton splitting
functions \kgamma are given in $x$-space in Ref.~\cite{Gluck:1983mm}. At leading order, $\kgamma^{(0)}$
contains only the quark splitting $k_q^{(0)}$. At higher orders, however, $\kgamma^{(1)}$
contains the splitting to gluons as well~\cite{Gluck:1991ee}.
For our NLO PDFs, the evolution includes the QCD splitting functions \pij up to two loops,
of order $\alphaSn{2}$, and the photon splitting functions up to $\kgamma^{(1)}$, of order $\alphaem\alphaS$.

We must briefly discuss the structure function of the photon here, which is defined
as convolutions of the coefficient functions with the respective quark and gluon
PDFs of the photon as well as the additional direct-photon coefficient function
\cgamma. In the \MSbar scheme,
\begin{equation}
  \cgamma(x) = 3 \left[ \left(x^2 + (1-x)^2\right) \ln\left(\frac{1-x}{x}\right) + 8x(1-x) - 1 \right] \, ,
  \label{eq:cgamma}
\end{equation}
which enters \ftwogamma weighted with $e_q^4$ for each quark flavour
and leads to significant negative contributions at large $x$, making
the decomposition of \ftwogamma unstable. In the past, an alternative
scheme \DISgamma~\cite{Gluck:1991ee} has been proposed to avoid this
instability, where this direct contribution is absorbed into the quark
PDFs.

This change of factorisation scheme, while leaving \ftwogamma unchanged,
has consequences for the evolution, as the splitting functions \kgamma
are also redefined:
\begin{align}
  k_{q,\DISgamma}^{(1)} &= k_{q,\MSbar}^{(1)} - e_q^2 P_{qq}^{(0)} \ast \cgamma \, ,
  \label{eq:kgamma_redef} \\
  k_{g,\DISgamma}^{(1)} &= k_{g,\MSbar}^{(1)} - 2 \sum_q e_q^2 P_{gq}^{(0)} \ast \cgamma \, .
  \label{eq:kgamma_redef2}
\end{align}
For our evolution, we use a modified version of the \Apfelpp DGLAP evolution
library~\cite{Bertone:2017gds} that we call \Apfelppp, which supports the
anomalous terms in photon-PDF evolution and is documented in
Ref.~\cite{Epping2023}. This public version of the code is
set up to perform the evolution in the \DISgamma scheme. We further modify this version
by redefining the non-singlet, singlet and gluon splitting functions of
Eqs.~(\ref{eq:kgamma_redef}) and~(\ref{eq:kgamma_redef2})
to reflect the \MSbar scheme, as the \DISgamma scheme is suited only to inclusive DIS
and not to the jet production considered here. The pure QCD splitting functions \pij do not
receive any corrections from this change of scheme.

Additionally, the evolution code uses the existing \Apfelpp machinery to generate heavier quarks
dynamically once the scale reaches their thresholds at or above \qzero.
By default, this results in the 5-flavour scheme; we modify the code to perform the full evolution
in the 4-flavour scheme.
The strong coupling runs at three loops with four flavours, with $\alphaS(m_Z) = 0.112$ (for
equivalence with 5-flavour running to the standard $\alphaS(m_Z) = 0.118$ value).

\subsection{PDF parametrisation}
\label{sec:param}

We use an NLO parametrisation for partons in the photon, proposed by
S{\l}omi{\'n}ski, Abramowicz and Levy (SAL)~\cite{Slominski:2005bw}.
We use the default starting scale of \Apfelppp, $\qzero =
\qty{1.3}{\GeV}$, as changing this does not significantly affect the
fit. As the evolution starts at the charm threshold, also set to
\qty{1.3}{\GeV}, we only parametrise the $u,d,s$ quarks and the gluon. The
charm quark is generated radiatively above its threshold.
In the 4-flavour scheme used throughout, there are
no bottom quarks in the photon. The parton contents are parametrised
at \qzero as follows:
\begin{itemize}
\item
All parametrised quarks are described as a sum
of charge-dependent, point-like contributions and hadronic
contributions:
\begin{equation}
    f_q^{(\gamma)}(x) = f_{\bar{q}}^{(\gamma)}(x) = e_q^2 \, A_{\mathrm{PL}} \, \frac{x^2 + (1-x)^2}{1 - B_\mathrm{PL}\ln(1-x)} + f_q^\mathrm{had}(x) \, ,
    \label{point-like}
\end{equation}
where \apl and \bpl are free parameters.

\item
$f_q^{\mathrm{had}}$ contains the hadronic term, parametrised for the $u,d$ quarks as
\begin{equation}
    f_u^\mathrm{had}(x) = f_d^\mathrm{had}(x) = A^\mathrm{had} x^{B^{\mathrm{had}}}(1-x)^{C^{\mathrm{had}}} \, ,
    \label{hadronic_structure_functions_ud}
\end{equation}
where \ahad and \bhad are free parameters.
As in SAL, the hadronic term is assumed to depend only on the quark mass,
so the difference between the $u,d$ quarks comes from the quark charge in the point-like term.

\item
The heavier strange quark is suppressed by a fixed factor $r_{s} = 0.3$, identical to SAL,
\begin{equation}
    f_s^\mathrm{had}(x) = r_{s} f_d^\mathrm{had}(x) \, .
    \label{hadronic_structure_functions_s}
\end{equation}

\item
The gluon contains only the hadronic term, and is described by
\begin{equation}
    f_G^{(\gamma)}(x) = A^\mathrm{had}_G \, x^{B^{\mathrm{had}}_G} (1-x)^{C^{\mathrm{had}}_G} \, ,
\end{equation}
where \ahadg and \bhadg float freely in the fit.
\end{itemize}
Following the approach taken by SAL~\cite{Slominski:2005bw}, we fix
the large-$x$ exponents to $\chad = 1$, $\chadg = 3$, and the strangeness suppression
factor to $r_{s} = 0.3$; we revisit the implications of this choice in
Sections~\ref{sec:pdftune} and~\ref{sec:further}.

\section{Event generation}
\label{sec:evgen}
We derive our predictions using a pre-release version of \Sherpa-3.1.0~\cite{Sherpa:2024mfk}, which provides NLO-accurate hadron-level predictions for photoproduction~\cite{Hoeche:2023gme,Meinzinger:2023xuf}.
One-loop amplitudes are provided by \OpenLoops\,2~\cite{Buccioni:2019sur} for all setups, internally using the \Collier~\cite{Denner:2016kdg} library. The tree-level matrix elements are provided by \Comix~\cite{Gleisberg:2008fv}
and \Amegic~\cite{Krauss:2001iv}. To regulate infrared divergences, an automated implementation~\cite{Gleisberg:2007md,Schonherr:2017qcj} of the Catani--Seymour subtraction~\cite{Catani:1996vz} is used and the regulated matrix elements are matched to the \CSS parton shower using \MCatNLO matching~\cite{Schumann:2007mg,Hoeche:2011fd}.
The photon flux is provided by the EPA interface that is implemented in \Sherpa~\cite{Hoeche:2023gme}
and is described by the Weizsäcker--Williams formula~\cite{vonWeizsacker:1934nji,Williams:1934ad,Budnev:1975poe,Frixione:1993yw,Schuler:1996qr}, see Eq.~(\ref{eq:photonflux}).
Fragmentation of partons into hadrons is modelled with the \Ahadic hadronisation model~\cite{Chahal:2022rid}, tuned to \LEP data~\cite{Iskauskas:2026rxi}.

Throughout, we work in a 4-flavour scheme, with the bottom mass
set to $m_b = \qty{4.92}{\GeV}$.
We set the renormalisation and factorisation scales to half the scalar sum of the
transverse momenta of the final-state partons, $\mu_F = \mu_R = H_T / 2$, and the resummation scale,
which sets the starting scale of the parton shower, to its default $\mu_Q = \mu_F$.

For the \HERA event generation, we use the \pdfforlhc set~\cite{PDF4LHCWorkingGroup:2022cjn} for the proton and take the strong coupling from it.
For \LEP, we use the same value, $\alphaS(m_Z) = 0.118$, with three-loop running.
This is the five-flavour value corresponding to the four-flavour coupling
used in the evolution of the photon PDF, cf.~Section~\ref{sec:theory}.
All other settings, for instance those of the beam remnants and their primordial transverse momentum,
are kept at their tuned defaults.

\subsection{\texorpdfstring{\HERA}{HERA} setup}

We generate photoproduction events at the \HERA beam energies,
$E_e = \qty{27.6}{\GeV}$ and $E_p = \qty{920}{\GeV}$ or \qty{820}{\GeV},
depending on the analysis. For most analyses, the direct and resolved
processes are
\begin{equation}
  \gamma p \to 2\,j\,@\,\text{NLO} \quad\text{and}\quad \gamma p \to b\bar{b}\,@\,\text{NLO} \, ,
\end{equation}
matched to the \CSS dipole shower with \MCatNLO~\cite{Hoeche:2011fd}.
Here, $j$ is any light parton in the event generation: $u,d,s,c,g$.
For the three-jet observables of \texttt{ZEUS\_2007\_I756660}~\cite{ZEUS:2007xwd},
the resolved and direct processes are generated with one more jet at NLO accuracy,
\begin{equation}
  \gamma p \to 3\,j\,@\,\text{NLO} \quad\text{and}\quad \gamma p \to b\bar{b}j\,@\,\text{NLO} \, ,
\end{equation}
and for the four-jet observables of the same analysis, four jets are generated at NLO,
\begin{equation}
  \gamma p \to 4\,j\,@\,\text{NLO} \quad\text{and}\quad \gamma p \to b\bar{b}jj\,@\,\text{NLO} \, .
\end{equation}

\subsection{\texorpdfstring{\LEP}{LEP} setup}

We generate events at the \LEP centre-of-mass energies of the analyses, $\sqrt{s} = \qty{198}{\GeV}$ for
\texttt{OPAL\_2003\_I611415} and \texttt{L3\_2004\_I661114} and \qty{206}{\GeV} for \texttt{OPAL\_2008\_I754316}.
For all analyses, we generate direct, single-resolved, and double-resolved events
\begin{equation}
  \gamma \gamma \to 2\,j\,@\,\text{NLO} \quad\text{and}\quad \gamma \gamma \to b\bar{b}\,@\,\text{NLO} \, ,
\end{equation}
matched to the \CSS dipole shower with \MCatNLO~\cite{Hoeche:2011fd}.
For the single-resolved process, we only generate the events in one beam orientation,
and double the cross-sections. This is sufficient because all \LEP observables
used here are symmetric under the exchange of the two beams.

\subsection{Multi-parton interactions}
\label{subsec:mpi_model}

\Sherpa simulates multi-parton interactions (MPI) with \Amisic~\cite{Sherpa:2024mfk}, which is based on the model of Sjöstrand and van Zijl~\cite{Sjostrand:1987su}.
MPI are only simulated if both incoming particles are resolved, i.e.\ described by PDFs, in which case the resolved photon is treated like a hadron.
At \HERA, this is the case for the resolved component and at \LEP for the double-resolved one, whereas direct and single-resolved events receive no MPI.

The secondary scatters are leading-order QCD $2\to2$ processes, computed with the PDFs of the incoming photons and protons.
They are generated in decreasing order of transverse momentum \pT, starting from the scale of the hard process.
Their mean number per event is set by the ratio of the integrated $2\to2$ cross-section above \pTmin to the non-diffractive cross-section $\sigma_\mathrm{ND}$.
The latter is taken from the parametrisations of Schuler and Sjöstrand~\cite{Schuler:1993wr,Schuler:1996en} and rescaled by a factor \sigmaNDnorm.
For the photon, it is a sum over $\rho$, $\omega$, $\phi$, and $J/\psi$ states in the vector-meson-dominance picture, weighted with their couplings to the photon.
The scatter probability also depends on the impact parameter $b$, through the overlap of the Gaussian matter distributions of the two incoming particles, whose radii are parameters of the model.
At each centre-of-mass energy, both radii are multiplied by a common factor such that the $b$-integrated probability for at least one scatter reproduces $\sigma_\mathrm{ND}$;
where the integrated $2\to2$ cross-section above \pTmin is smaller than $\sigma_\mathrm{ND}$, the radii are used unchanged.
In events with a hard process, $b$ is drawn from this overlap, favouring central collisions with more MPI activity.
The model does not separate the vector-meson-like and the anomalous, point-like components of the resolved photon: the full photon PDF enters the secondary scatters, and the non-diffractive cross-section and the matter distribution are those of the vector-meson picture.

The divergence of the $2\to2$ matrix elements at small \pT is regularised by a factor $\left[\pT^2/(\pT^2+\pTnought^2)\right]^2$.
The strong coupling is evaluated at a scale proportional to $\pT^2+\pTnought^2$ and
no scatters are generated below the cut-off \pTmin.
Both scales are parametrised in terms of the centre-of-mass energy \Ecms of the photon--proton or photon--photon system as
\begin{align}
  \pTnought &= \max\left[\pTnoughtIR,\ \pTnoughtref \left(\frac{\Ecms}{\Ecmsref}\right)^{2\eta}\right] \, ,
  \label{eq:IR_regulator} \\
  \pTmin &= \max\left[\pTminIR,\ \pTminref \left(\frac{\Ecms}{\Ecmsref}\right)^{2\eta}\right] \, ,
  \label{eq:IR_cutoff}
\end{align}
with a common energy-scaling exponent $\eta$, a reference energy \Ecmsref, and lower bounds \pTnoughtIR and \pTminIR, by default \qty{0.5}{\GeV} and \qty{1}{\GeV}.
We set \Ecmsref to the beam centre-of-mass energy of the respective collider.
Since the photon energies follow the EPA spectrum, \Ecms varies from event to event.

\section{Fitting methodology}\label{sec:method}

The fit follows the surrogate-based strategy outlined in
Section~\ref{sec:plfit}; here we describe its implementation.

\subsection{Surrogate optimisation, or ``tuning''}
\label{subsec:surrogate_tuning}

In the surrogate approach, every bin of every histogram is given an
independent surrogate function of the continuous parameters being
optimised, typically an interpolation between randomly sampled points
within the parameter-space hypercube or hypersphere.

The \Professor package uses simple multivariate polynomials, within
possibly transformed parameter measures; more advanced emulator
techniques such as neural networks or Gaussian processes equally meet
the criteria. Armed with an ensemble of surrogate functions,
applications such as serial optimisation become tractable, even in
real-time for visualisation tasks.
In this study, we use \Professor v2.5.9 with third-order polynomials for the PDF fit.
The PDF fit has six free parameters, and the MPI tunes have six for
\HERA and five for \LEP, where there is no proton radius.

The \kbd{prof2-sample} program was used to sample parameter points
$\{\pvec\}$ uniformly within the hypercube of PDF parameters
documented in Section~\ref{sec:param}, with the ranges listed in
Table~\ref{tab:plg26-params}, injecting those parameters into
steering scripts for a custom PDF grid-writer program linked against
the \Apfelppp evolution library described in
Section~\ref{sec:theory}. The grid-writer evaluated the functional
forms from Section~\ref{sec:param} to set the PDF distributions at the
input scale $\qzero = \qty{1.3}{\GeV}$, and the \Apfelppp evolution
was used to generate a PDF grid in $(x, Q^2)$ in the \LHAPDF\,6 data
format~\cite{Buckley:2014ana}, from $x = \numrange{1e-5}{1}$, and in
$Q^2$ up to $\qty{1e4}{\GeV\squared}$.  Parameter points whose evolved
PDFs became negative were rejected, so that they entered neither the
MC reweighting nor, consequently, the interpolation. The remaining PDF
variations were combined into a single \LHAPDF set, which was provided
to the \Sherpa MC generator as if it were a standard, established
global PDF with fit-uncertainty variations to be propagated via
event-weighting.

A single set of runs to cover the involved process types was then
sufficient to propagate the randomly sampled PDF parametrisations
through \Sherpa, using its existing on-the-fly reweighting
implementation at NLO~\cite{Bothmann:2016nao} to predict the
differential cross-sections for all input PDFs at once, via the set of photoproduction
event-analyses implemented in the \Rivet analysis
framework. The Python interface to the \yoda histogramming
package~\cite{Buckley:2023xqh} used by \rivet was used to split the
resulting nominal and variation observable-histograms into
per-variation copies associated with the original sampled parameter
variations, and used as input to the standard \Professor
interpolation-building and tuning.

The \Professor tuning finds an optimised parameter vector \pvectune
that minimises the heuristic goodness-of-fit function
\begin{equation}
  \chi^2 = \sum_b \frac{(y_b(\pvec) - d_b)^2}{\sigma^2(y_b(\pvec)) + \sigma^2(d_b) + (\epsilon_b d_b)^2} \, ,
  \label{eq:prof_chi2}
\end{equation}
where $b$ indexes the set of summed-over observable bins, $y_b(\pvec)$
is the \Professor surrogate prediction for bin $b$ with parameters
$\pvec$, $d_b$ is the data value of bin $b$, and the denominator
contains a quadrature sum of uncertainties for bin $b$. Of these,
$\sigma(y_b(\pvec))$ is a representative uncertainty from the
surrogates --- in this case the median statistical uncertainty of the
MC runs --- and $\sigma(d_b)$ and $\epsilon_b d_b$ are respectively
the reference-data uncertainty and an \emph{ad hoc} uncertainty for MC
theory; we use a value of 5\,\% for the relative uncertainty
$\epsilon_b$ on all bins. This form of the fit-function treats all
bins as independent, which is unavoidable as the available data lacks
the uncertainty breakdowns and correlations required for a more
precise maximum-likelihood fit.

To avoid local minima, we additionally scanned \num{10000} points of the
parameter space with the \Professor $n$-ball sampler,\footnote{This
feature restricts sample selection to the interior of the hypersphere
contained within the hypercube of normalised parameter ranges,
focusing on the central region of the volume to avoid points with at
least one parameter near its limits, which dominate as dimensionality
increases.} and used the point with the lowest $\chi^2$ as the starting
point of the \Minuit minimisation.

As, due to model and correlation-data shortcomings, the fit function
is not in practice distributed like a classical $\chi^2$
test-statistic, the replica method is used to estimate fit
uncertainties resulting from the observable-measurement
uncertainties. To produce the replicas, we start from the reference
data in the \yoda files and create replica datasets by shifting the
data value $d_b$ of each bin $b$ in replica $r$ as
\begin{equation}
  d_b^{(r)} = d_b + \sigma_b \, \mc{N}^{(r)}(0,1) \, ,
  \label{eq:gaussian_smearing}
\end{equation}
where $\sigma_b$ is the symmetric Gaussian uncertainty of bin $b$ and $\mc{N}^{(r)}(0,1)$ is a
standard-normal random number. Asymmetric uncertainties are symmetrised for the smearing,
but stay asymmetric in the replica, and the uncertainties are not smeared, $\sigma^{(r)}_b = \sigma_b$.
Within a replica, the random numbers are generated from a single fixed seed.

The PDF fit was repeated for each of the 100 replica datasets, resulting in 100 replica PDFs.
The central PDF is the fit to the unsmeared data, and the replica PDFs can deviate from it
in either direction, so that the resulting uncertainties are asymmetric.
The uncertainty bands in Section~\ref{sec:results} show the 68\,\% interval of the replica PDFs.
The replica uncertainties therefore reflect the experimental uncertainties of the individual bins.
They do not account for correlations between systematic uncertainties, or for theory variations
such as the choice of scales.

\subsection{MPI tuning}
\label{subsec:mpi_tuning}

A complicating factor in this analysis is that photoproduction has
significant phase-space overlap with the event-generator mechanism of
multi-parton interactions (MPI), which model secondary partonic
scatterings within hadron-hadron collisions to describe sensitive $ep$
and hadron-collider observables. MPI models are highly
phenomenological rather than first-principles, and the setting of MPI
parameters has long been a core focus of MC tuning.

As the resolved photon is an effective hadron, and one expected to
have quite different properties from the (anti)proton collisions on
which the \Sherpa default MPI tune was established, poor
MPI modelling of resolved $\gamma p$ and $\gamma\gamma$ collisions risks incorrect conclusions about
resolved-photon structure: effects that should be described by the MPI
model would be ``absorbed'' into the PDFs. Moreover, as \Sherpa-3.1 has
an untuned MPI model for \HERA and \LEP photoproduction, setting a good
baseline description for these scenarios with existing PDFs is useful
in its own right.

We therefore performed the tuning in two steps. First, we tune the MPI
parameters listed in Table~\ref{tab:mpi_results} with the photon PDF
fixed to the historical SaS\,1M
set~\cite{Schuler:1995fk,Schuler:1996fc}, the \Sherpa default. We then
fix the MPI parameters at their tuned values and perform a joint PDF
fit to the \HERA and \LEP data.  Both steps use largely the same
measurements: the MPI tunes
include additional distributions and extended ranges at lower
transverse energies, while the PDF fit additionally includes
\texttt{ZEUS\_1997\_I450085} and the shapes of the four-jet
distributions of \texttt{ZEUS\_2007\_I756660}, given in detail in
Appendices~\ref{sec:AppMPI} and~\ref{sec:AppPDF}.

Separate MPI tunes were made for $\gamma p$ collisions at \HERA and
$\gamma\gamma$ collisions at \LEP, both with SaS\,1M, and each
tune is used for the predictions of its collider in the PDF fit.  Due to the
different beam energies and particles at the two colliders, these need
to be tuned separately.  It may be possible to do a combined tune, but
this is an active field of research (see
Ref.~\cite{Butterworth:2024hvb}) and for the purposes of this
exercise, keeping the complexity to a minimum with simple individual
tunes was preferred.  With \HERA, the additional proton-radius
parameter must also be considered; here, we use single Gaussians to
model the spatial form-factors of each particle.

For \HERA, 200 sets of parameters were used for the MPI variations, and
with \LEP, reflecting the one less free parameter, we used 150. For
both, we performed the interpolation with second-order polynomials.

\subsection{Observable selection}

The event analysis was performed using \rivet[4] and \yoda[2], with
the analyses listed in Table~\ref{tab:rivet_analyses}.  We exclude the
distributions and bins restricted to the direct-enriched region, as
including them changes neither the MPI nor PDF fit.  We also exclude
the four-jet distributions of \texttt{ZEUS\_2007\_I756660} from
the MPI tunes, and use them in the PDF fit only as shapes, normalised
to unit area, as the NLO predictions lie consistently above the data
and would otherwise pull the fit.  The \xgammaobs cuts keep bins
that straddle the cut, e.g.\ $0.7 < \xgammaobs < 0.85$ in
\texttt{H1\_2002\_I581409}, so the resolved-enriched selections extend
into the direct-enriched region.  We selected the data by hand,
choosing the histograms and bins that are most sensitive to the tuned
parameters, mostly from envelope plots, which compare the spread of
the predictions for all sampled parameter points with the
data. Automated filters, such as the envelope filter of \Professor,
serve only as an additional safeguard.

\begin{table}
    \centering
    \begin{tabular}{lcl}\toprule
      \rivet tag &  Ref. & Observables \\\midrule
      \texttt{H1\_2002\_I581409} & \cite{H1:2002apm} & Dijet cross-sections in photoproduction \\
      \texttt{H1\_2006\_I711847} & \cite{H1:2006rre} & Photoproduction of dijets with high transverse momenta \\
      \texttt{ZEUS\_1997\_I450085} & \cite{ZEUS:1997fwn} & Dijet cross-sections in photoproduction \\
      \texttt{ZEUS\_2001\_I568665} & \cite{ZEUS:2001zoq} & Dijet cross-sections in photoproduction \\
      \texttt{ZEUS\_2007\_I753991} & \cite{ZEUS:2007njl} & High-\ET dijet photoproduction \\
      \texttt{ZEUS\_2007\_I756660} & \cite{ZEUS:2007xwd} & Three- and four-jet final states in photoproduction \\
      \texttt{ZEUS\_2012\_I1116258} & \cite{ZEUS:2012pcn} & Inclusive-jet photoproduction \\
      \texttt{L3\_2004\_I661114} & \cite{L3:2004ehh} & Inclusive jet production in two-photon collisions \\
      \texttt{OPAL\_2003\_I611415} & \cite{OPAL:2003hoh} & Dijet production in photon-photon collisions \\
      \texttt{OPAL\_2008\_I754316} & \cite{OPAL:2007jeb} & Inclusive jet production in photon-photon collisions \\ \bottomrule
    \end{tabular}
    \caption{Data used in both PDF fitting and MPI tuning.
      The selected histograms
      and bins are listed in Appendices~\ref{sec:AppMPI} and~\ref{sec:AppPDF}.}
    \label{tab:rivet_analyses}
\end{table}

\subsection{PDF sampling}
\label{subsec:pdf_sampling}

The six PDF parameters were sampled as described in
Section~\ref{subsec:surrogate_tuning}, within the ranges
listed in Table~\ref{tab:plg26-params}, using 3000 PDFs at
once for all PDF tuning MC generator runs.

\section{Fit results}
\label{sec:results}

\subsection{MPI tunes}
\label{subsec:mpi_results}

\begin{table}[btp]
\centering
\begin{tabular}{lccc}
\toprule
Parameter & Range & \HERA & \LEP \\
\midrule
\pTnoughtref [\unit{\GeV}] & $[0.25, 10]$ & $1.77$ & $5.9$ \\
\pTminref [\unit{\GeV}] & $[0.8, 3]$ & $1.683$ & $1.56$ \\
$\eta$ & $[0, 0.3]$ & $0.2637$ & $0.055$ \\
\sigmaNDnorm & $[0.1, 1.2]$ & $1.069$ & $0.103$ \\
$R_\gamma$ [\unit{\femto\metre}] & $[0.6, 0.9]$ & $0.6743$ & $0.6172$ \\
$R_p$ [\unit{\femto\metre}] & $[0.6, 1.1]$ & $1.0496$ & --- \\
\midrule
$\chi^2/N_\mathrm{df}$ & & $1698.2/508 = 3.34$ & $498.2/129 = 3.86$ \\
\bottomrule
\end{tabular}
\caption{Results for both \HERA and \LEP MPI tunes, as used for the final PDF tune.}
\label{tab:mpi_results}
\end{table}

Table~\ref{tab:mpi_results} lists the tuned MPI parameters for \HERA and \LEP,
and Table~\ref{tab:mpi_chi2} documents the fit quality of each analysis.
The tuned predictions give $\chi^2/N_\mathrm{df} = 3.34$ for \HERA and $3.86$ for \LEP.

At \HERA, all parameters lie inside their ranges, with $\eta$ and $R_p$ close to the upper edges.
The reference scales \pTnoughtref and \pTminref apply at \Ecmsref, the centre-of-mass energy
of the collider, and decrease towards lower photon--proton energies,
cf.\ Eqs.~(\ref{eq:IR_regulator}) and~(\ref{eq:IR_cutoff}):
at $\Ecms = \qty{200}{\GeV}$, the tuned values correspond to $\pTnought \approx \qty{1.4}{\GeV}$
and $\pTmin \approx \qty{1.3}{\GeV}$.

At \LEP, MPI enter only the double-resolved component and
the tune prefers a larger regulator than at \HERA, $\pTnought \approx \qty{4.8}{\GeV}$ at
$\Ecms = \qty{30}{\GeV}$, while \pTmin is similar at \qty{1.3}{\GeV}, and \sigmaNDnorm lies at the
lower edge of its range.
The photon radius, $R_\gamma = \qty{0.62}{\femto\metre}$, close to the lower edge of its range,
is similar to that of the \HERA tune, \qty{0.67}{\femto\metre}. For both colliders, the fits prefer
the photon to be smaller than the \HERA proton, as expected.

Per analysis, the H1 dijet data are described well, with $\chi^2/N_\mathrm{bins} \approx 1$.
At \HERA, the ZEUS inclusive-jet and three-jet data, \texttt{ZEUS\_2012\_I1116258} and
\texttt{ZEUS\_2007\_I756660}, account for two thirds of the $\chi^2$, and at \LEP,
\texttt{OPAL\_2003\_I611415} contributes 411.7 of 498.2.
At \LEP, \texttt{OPAL\_2008\_I754316} has the largest $\chi^2$ per bin, 6.24, from 11 bins.
\texttt{ZEUS\_2012\_I1116258} and \texttt{OPAL\_2008\_I754316} are poorly modelled across the
full range of distributions. In \texttt{ZEUS\_2012\_I1116258}, low $E_\mathrm{T}$ and most
forward $\eta$ bins affect the $\chi^2$ the most, while in \texttt{OPAL\_2008\_I754316}, intermediate
  $p_\mathrm{T}$ bins between \qty{7.5}{\GeV} and \qty{15}{\GeV} in both $\eta$ regions are the
worst described.

\begin{figure}[htbp]
    \centering
    \subfloat[ZEUS three-jet photoproduction~\cite{ZEUS:2007xwd}: \xgammaobs for $25 < M_{3j} < \qty{50}{\GeV}$, jets with $E_\mathrm{T} > \qty{6}{\GeV}$ and $\vert\eta\vert < 2.4$]{
        \includegraphics[width=0.40\columnwidth]{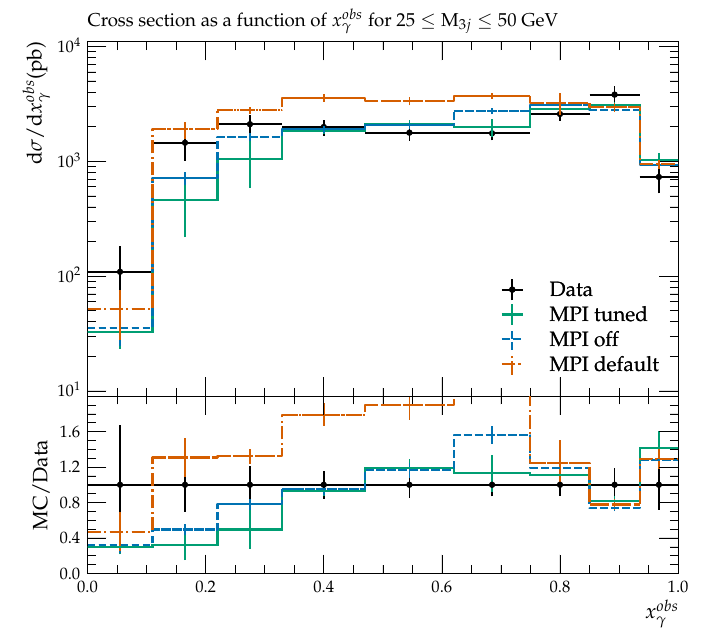}
        \label{fig:mpi_tune_zeus07_xgamma}
    }
    \hfill
    \subfloat[ZEUS three-jet photoproduction~\cite{ZEUS:2007xwd}: $\cos\psi_3$ for $M_{3j} > \qty{50}{\GeV}$, jets with $E_\mathrm{T} > \qty{6}{\GeV}$ and $\vert\eta\vert < 2.4$]{
        \includegraphics[width=0.40\columnwidth]{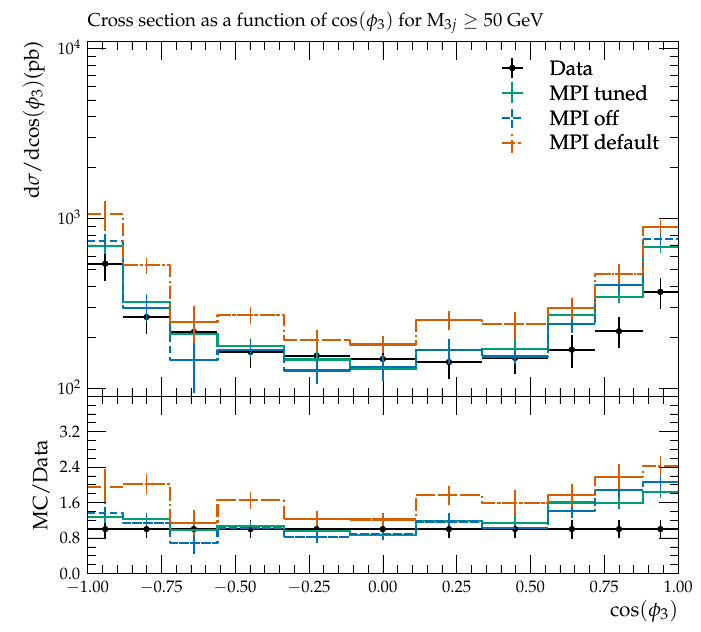}
        \label{fig:mpi_tune_zeus07_psi3}
    }

    \vspace{0.3cm}

    \subfloat[ZEUS inclusive-jet photoproduction~\cite{ZEUS:2012pcn}: $\eta^\mathrm{jet}$ of $k_t$ jets with $E_\mathrm{T} > \qty{21}{\GeV}$]{
        \includegraphics[width=0.40\columnwidth]{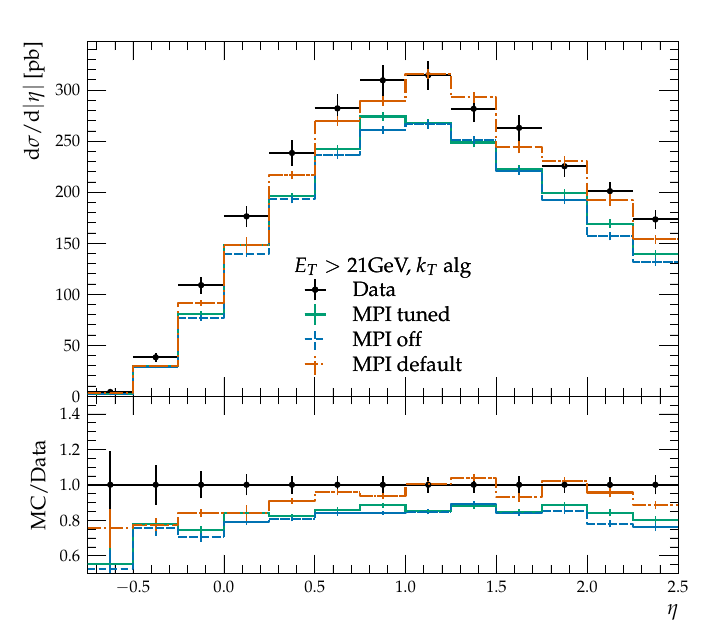}
        \label{fig:mpi_tune_zeus12_eta}
    }
    \hfill
    \subfloat[L3 inclusive jets in $\gamma\gamma$ collisions~\cite{L3:2004ehh}: $p_\mathrm{T}^\mathrm{jet}$ for $\vert\eta^\mathrm{jet}\vert < 1$]{
        \includegraphics[width=0.40\columnwidth]{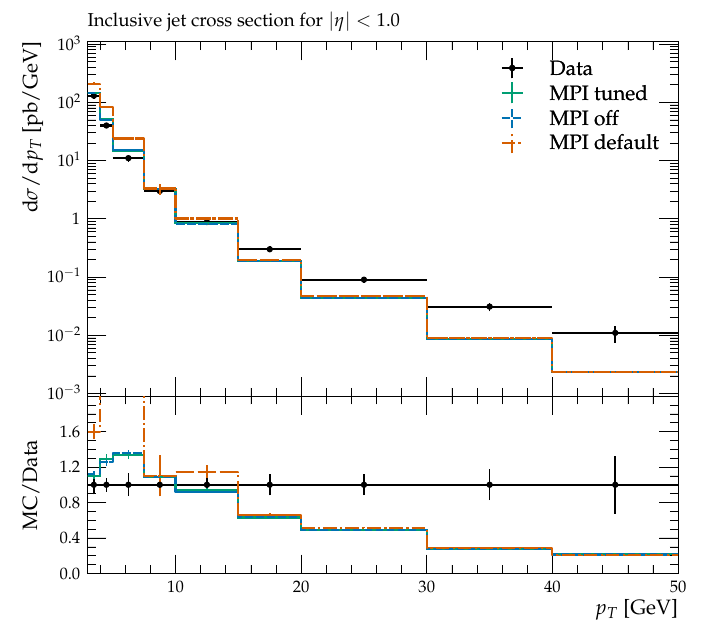}
        \label{fig:mpi_tune_l3_pt}
    }

    \vspace{0.3cm}

    \subfloat[OPAL dijets in $\gamma\gamma$ collisions~\cite{OPAL:2003hoh}: \xgammaobs for $5 < \bar{E}_\mathrm{T} < \qty{7}{\GeV}$, jets with $\vert\eta\vert < 2$]{
        \includegraphics[width=0.40\columnwidth]{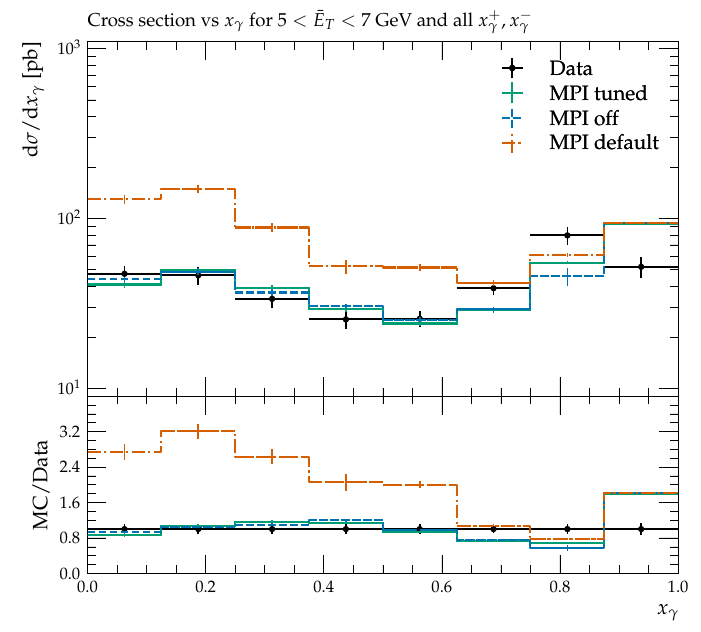}
        \label{fig:mpi_tune_opal_all}
    }
    \hfill
    \subfloat[OPAL dijets in $\gamma\gamma$ collisions~\cite{OPAL:2003hoh}: as in (e), with both $x_\gamma^+$ and $x_\gamma^- < 0.75$]{
        \includegraphics[width=0.40\columnwidth]{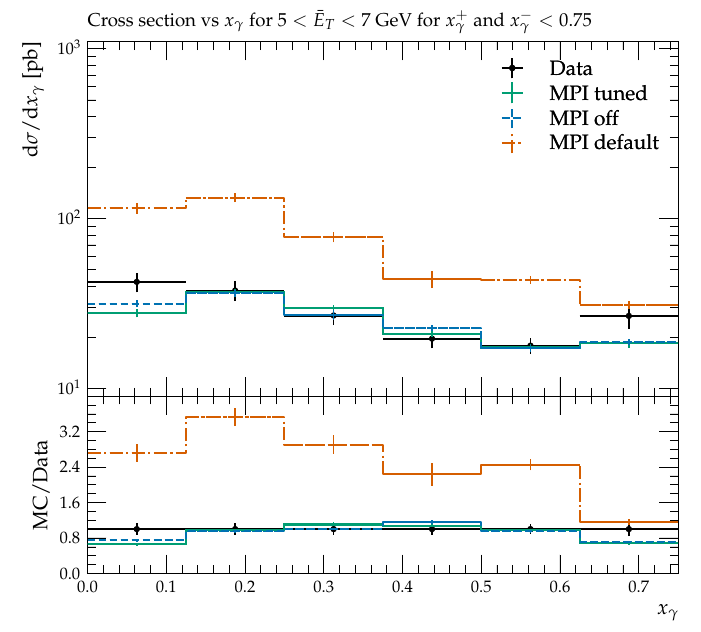}
        \label{fig:mpi_tune_opal_dr}
    }

    \caption{Selected observable distributions from the MPI tunes: data compared with \Sherpa predictions with the
      tuned MPI parameters of Table~\ref{tab:mpi_results}, without MPI, and with the default MPI
      parameters, all with SaS\,1M. The \HERA data are in photoproduction, $Q^2 < \qty{1}{\GeV\squared}$;
      all jets are defined with the $k_t$ algorithm. The lower panels show the ratios to the data.}
    \label{fig:mpi_tune}
\end{figure}

Figure~\ref{fig:mpi_tune} compares the tuned predictions with data for a selection of the
distributions in the tune, together with predictions without MPI and with the default MPI
parameters of \Sherpa.
At \LEP, the tuned predictions are close to those without MPI, consistent with \sigmaNDnorm at
the lower edge of its range, whereas the default parameters overshoot the OPAL data at low
\xgammaobs by more than a factor of two.
Other than for the ZEUS inclusive-jet photoproduction data~\cite{ZEUS:2012pcn},
the tuning is seen to significantly improve the data description, with the tunes suppressing
MPI activity to a level closer to the MPI-off mode.

\begin{table}[htbp]
\centering
\begin{tabular}{lrrr}
\toprule
Analysis & Bins & $\chi^2$ & $\chi^2$/bin \\
\midrule
\multicolumn{4}{l}{\HERA}\\
\texttt{H1\_2002\_I581409} & 67 & 73.2 & 1.09 \\
\texttt{H1\_2006\_I711847} & 43 & 37.7 & 0.88 \\
\texttt{ZEUS\_2001\_I568665} & 56 & 216.1 & 3.86 \\
\texttt{ZEUS\_2007\_I753991} & 62 & 240.7 & 3.88 \\
\texttt{ZEUS\_2007\_I756660} & 192 & 341.4 & 1.78 \\
\texttt{ZEUS\_2012\_I1116258} & 94 & 788.9 & 8.40 \\
\midrule
\multicolumn{4}{l}{\LEP}\\
\texttt{L3\_2004\_I661114} & 5 & 17.8 & 3.56 \\
\texttt{OPAL\_2003\_I611415} & 113 & 411.7 & 3.64 \\
\texttt{OPAL\_2008\_I754316} & 11 & 68.6 & 6.24 \\
\midrule
Total \HERA & 514 & 1698.2 & 3.34 \\
Total \LEP & 129 & 498.2 & 3.86 \\
\bottomrule
\end{tabular}
\caption{The fit quality, $\chi^2/N_\mathrm{bins}$ broken down by analysis for the MPI tunes.
These are calculated against the bins used in the fit, tabulated in Appendix~\ref{sec:AppMPI}.}
\label{tab:mpi_chi2}
\end{table}

\clearpage

\subsection{PDF tune}
\label{sec:pdftune}

The PDF tune was performed using the parameter ranges given in
Table~\ref{tab:plg26-params}, with the MPI modelled using the SaS\,1M
PDF and tuned parameters, and the sampled PDFs for the resolved
scattering process. The MC statistics were doubled relative to the MPI
tunes for analyses where the median
$\sigma_\mathrm{MC}/\sigma_\mathrm{data}$ was greater than 0.50
(detailed in Appendix~\ref{sec:AppMPI}, Table~\ref{tab:mpi_mcstat}).

The final fit metrics are given
in Table~\ref{tab:plg26-chi2-truth}, with significant improvements
seen in the ZEUS dijet and inclusive-jet data compared with the MPI tunes with SaS\,1M
(Table~\ref{tab:mpi_chi2}): $\chi^2/N_\mathrm{bins}$ falls from 3.9 to 0.8 and 1.3 for the dijet data,
\texttt{ZEUS\_2001\_I568665} and \texttt{ZEUS\_2007\_I753991}, and from 8.4 to 1.9 for the
inclusive-jet data, \texttt{ZEUS\_2012\_I1116258}.
In contrast, the ZEUS three-jet data, \texttt{ZEUS\_2007\_I756660}, and the L3 inclusive-jet and
OPAL dijet data, \texttt{L3\_2004\_I661114} and \texttt{OPAL\_2003\_I611415}, are described less well
than in the MPI tunes, and the three-jet and OPAL dijet data together account for three quarters
of the total $\chi^2$. As the bin selections of the PDF fit and the MPI tunes differ, these
comparisons are indicative only.

\begin{figure}[ptb]
    \centering

    \subfloat[ZEUS high-\ET dijet photoproduction~\cite{ZEUS:2007njl}: \xgammaobs]{
        \includegraphics[width=0.42\columnwidth]{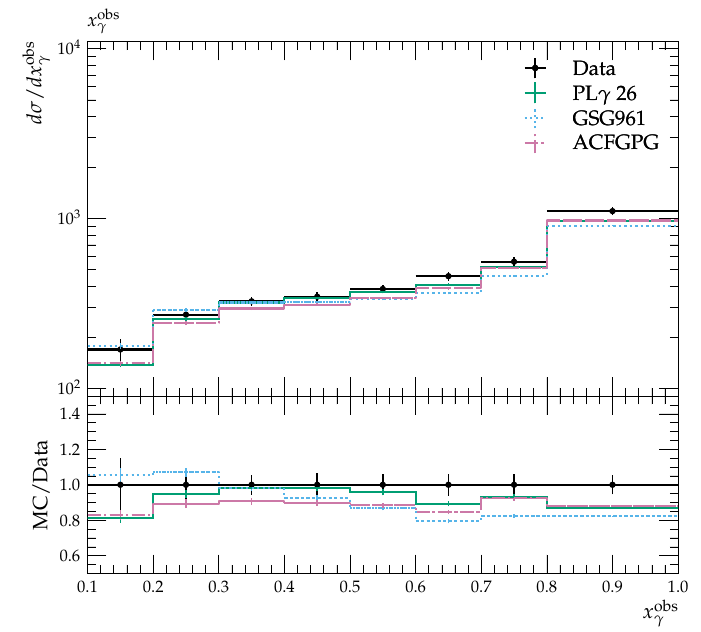}
        \label{fig:finalobs_d19-x01-y01_z2007_753991}
    }
    \hfill
    \subfloat[ZEUS inclusive-jet photoproduction~\cite{ZEUS:2012pcn}: $\eta^\mathrm{jet}$ of jets with $E_\mathrm{T} > \qty{21}{\GeV}$]{
        \includegraphics[width=0.42\columnwidth]{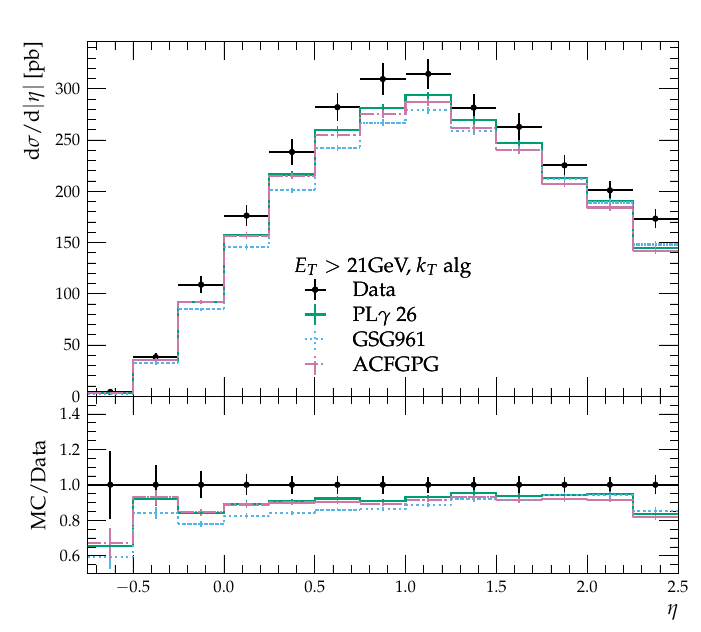}
        \label{fig:finalobs_d03-x01-y01_z2012}
    }

    \subfloat[ZEUS dijet photoproduction~\cite{ZEUS:2001zoq}: $\eta^\mathrm{jet2}$ for $\xgammaobs < 0.75$ and $0 < \eta^\mathrm{jet1} < 1$]{
        \includegraphics[width=0.42\columnwidth]{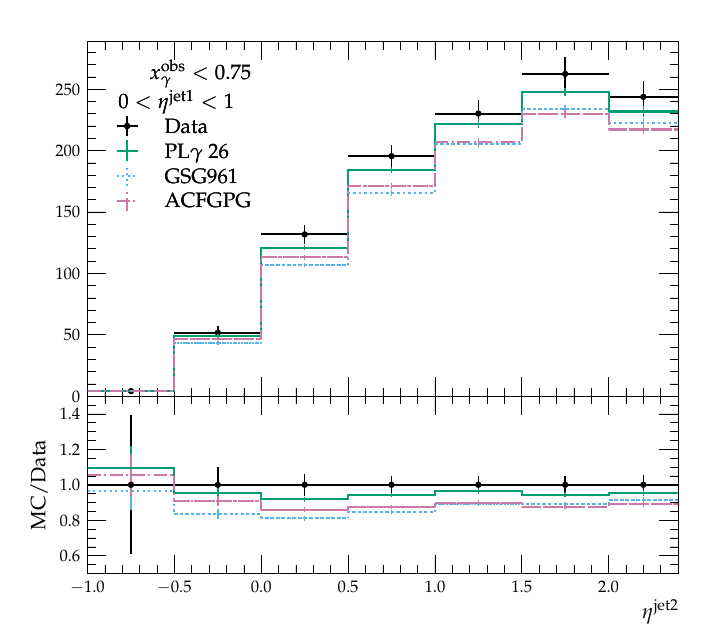}
        \label{fig:finalobs_d18-x01-y01_z2001}
    }
    \hfill
    \subfloat[ZEUS dijet photoproduction~\cite{ZEUS:2001zoq}: as in (c), for $1 < \eta^\mathrm{jet1} < 2.4$]{
        \includegraphics[width=0.42\columnwidth]{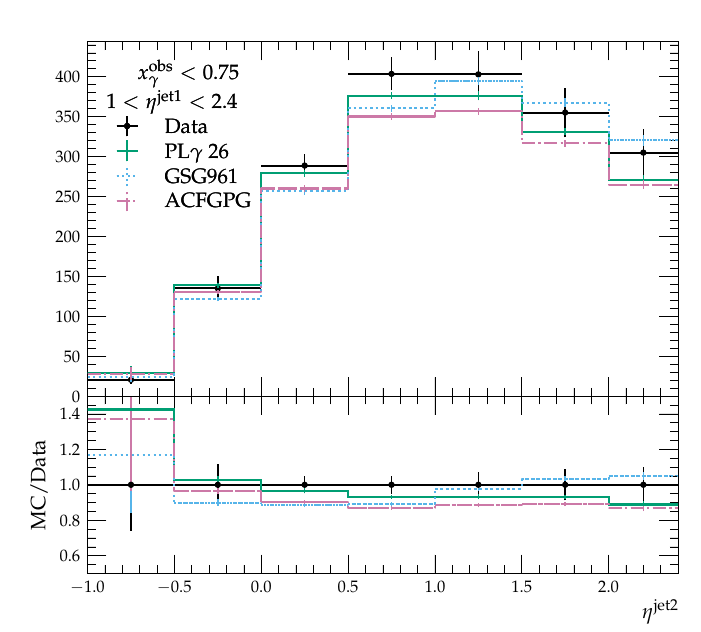}
        \label{fig:finalobs_d19-x01-y01_z2001}
    }

    \subfloat[L3 inclusive jets in $\gamma\gamma$ collisions~\cite{L3:2004ehh}: $p_\mathrm{T}^\mathrm{jet}$ for $\vert\eta^\mathrm{jet}\vert < 1$]{
        \includegraphics[width=0.42\columnwidth]{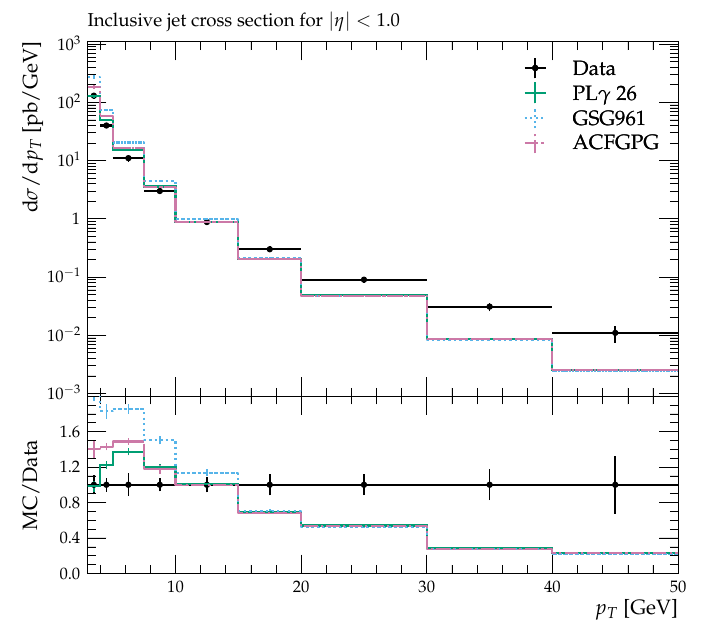}
        \label{fig:finalobs_d01-x01-y01_L3}
    }
    \hfill
    \subfloat[OPAL dijets in $\gamma\gamma$ collisions~\cite{OPAL:2003hoh}: $\max\vert\eta^\mathrm{jet}\vert$ for $5 < \bar{E}_\mathrm{T} < \qty{7}{\GeV}$, with one $x_\gamma^\pm < 0.75$]{
        \includegraphics[width=0.42\columnwidth]{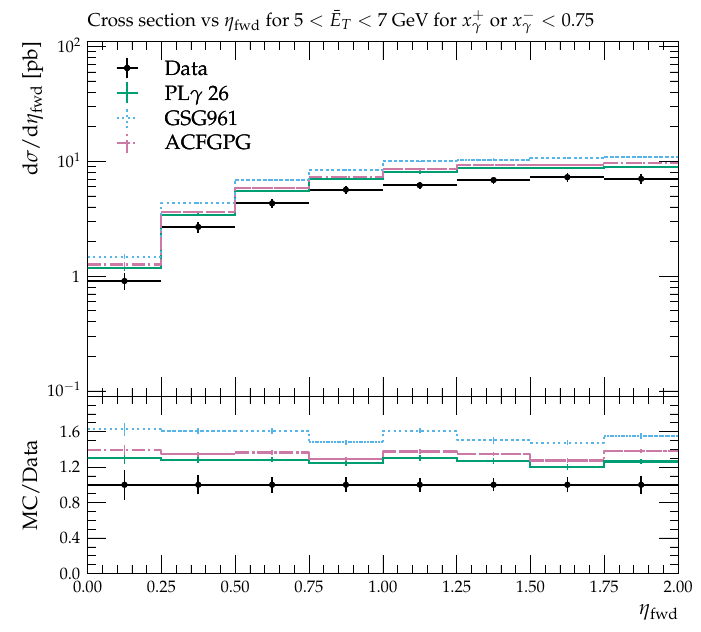}
        \label{fig:finalobs_d20-x01-y01_opal2003}
    }

    \caption{Selected observable distributions with final MPI tunes
      and PDF fits: \HERA and \LEP data compared with \Sherpa
      predictions using the tuned MPI configurations, and the
      resolved-photon PDFs \plgamma, GSG961 and ACFGPG.
      The \HERA data are in photoproduction, $Q^2 < \qty{1}{\GeV\squared}$;
      all jets are defined with the $k_t$ algorithm.
      The lower panels show the ratios to the data.}
    \label{fig:finalobs}
\end{figure}

Figure~\ref{fig:finalobs} compares the data for a selection of distributions with the predictions for
\plgamma and two older sets, GS96 (GSG961)~\cite{Gordon:1996pm} and ACFGP (ACFGPG)~\cite{Aurenche:1992sb},
all with the tuned MPI parameters.
At \HERA, \plgamma describes the resolved-enriched ZEUS dijet data best: its predictions lie within
5\,\% of the data for $0.2 < \xgammaobs < 0.6$, and within 8\,\% for $\xgammaobs < 0.75$ with a central
leading jet, whereas those with GS96 and ACFGP deviate by up to 20\,\%. For the ZEUS
inclusive-jet data, \plgamma and ACFGP agree to within a few per cent, both about 10\,\% below the data.
At \LEP, all three sets overshoot the OPAL dijet data with one $x_\gamma^\pm < 0.75$, \plgamma by
20--30\,\%, ACFGP by 30--40\,\% and GS96 by 50--60\,\%. All three also overshoot the L3 inclusive-jet
spectrum below $p_\mathrm{T} = \qty{10}{\GeV}$, \plgamma by up to 40\,\%, and undershoot it above
\qty{15}{\GeV}, independently of the PDF, by a factor of four at the highest $p_\mathrm{T}$.

\begin{table}[bp]
\centering
\begin{tabular}{lrrr}
\toprule
Analysis & Bins & $\chi^2$ & $\chi^2$/bin \\
\midrule
\multicolumn{4}{l}{\HERA}\\
\texttt{H1\_2002\_I581409} & 52 & 29.2 & 0.56 \\
\texttt{H1\_2006\_I711847} & 46 & 68.3 & 1.48 \\
\texttt{ZEUS\_1997\_I450085} & 29 & 17.1 & 0.59 \\
\texttt{ZEUS\_2001\_I568665} & 55 & 42.1 & 0.77 \\
\texttt{ZEUS\_2007\_I753991} & 58 & 75.6 & 1.30 \\
\texttt{ZEUS\_2007\_I756660} (3-jet) & 190 & 946.5 & 4.98 \\
\textcolor{gray}{\texttt{ZEUS\_2007\_I756660} (4-jet)} & \textcolor{gray}{183} & \textcolor{gray}{3639.6} & \textcolor{gray}{19.89} \\
\texttt{ZEUS\_2012\_I1116258} & 96 & 183.7 & 1.91 \\
\midrule
\multicolumn{4}{l}{\LEP}\\
\texttt{L3\_2004\_I661114} & 9 & 54.8 & 6.09 \\
\texttt{OPAL\_2003\_I611415} & 112 & 507.3 & 4.53 \\
\texttt{OPAL\_2008\_I754316} & 12 & 32.9 & 2.74 \\
\midrule
Total & 659 & 1957.6 & 2.97 \\
\bottomrule
\end{tabular}
\par\smallskip
\caption{Calculated $\chi^2$s of \plgamma PDF fit.
The reported $\chi^2$ is calculated using the formula
in Eq.~(\ref{eq:prof_chi2}).
The four-jet distributions of \texttt{ZEUS\_2007\_I756660} (grey) are used only for a shape-based
fit, normalised to unit area (\Professor's \texttt{normA} mode).
These four-jet $\chi^2$ values are not included in the total.}
\label{tab:plg26-chi2-truth}
\end{table}

\begin{table}[btp]
\centering
\begin{tabular}{lrrrc}
\toprule
Parameter & Central & Mean & Std.~dev. & Sampling range \\
\midrule
\apl &  $7.991$ & $5.967$ & $1.505$ & $[0, 8]$ \\
\bpl & $2.378$ & $1.700$ & $0.464$ & $[0, 3.5]$ \\
\ahad & $0.5985$ & $0.5383$ & $0.1760$ & $[0, 0.6]$ \\
\bhad & $1.4670$ & $1.1274$ & $0.4047$ & $[-2.5, 1.5]$ \\
\ahadg & $0.0001$ & $0.6263$ & $0.6523$ & $[0, 2]$ \\
\bhadg & $-0.8604$ & $0.4556$ & $0.7518$ & $[-3, 1]$ \\
\bottomrule
\end{tabular}
\par\smallskip
\caption{Fitted parameters of \plgamma: central tune and replica statistics (100 replicas). The mean
and standard deviation are calculated from the parameter values of the 100 replicas, as well as the central tune. While
the sampling ranges extend quite wide, the fit was restricted to the sub-box $\bhad\!, \bhadg > -2.0$. This
avoids a divergent momentum integral when all $x$ is considered. This also means that some replicas that would prefer less
physical \bhad, \bhadg values are pinned against this boundary.}
\label{tab:plg26-params}
\end{table}

\begin{figure}[ptb]
    \centering

    \subfloat[\qsquared = \qty{2.0}{\GeV \squared}]{
        \includegraphics[width=0.48\columnwidth]{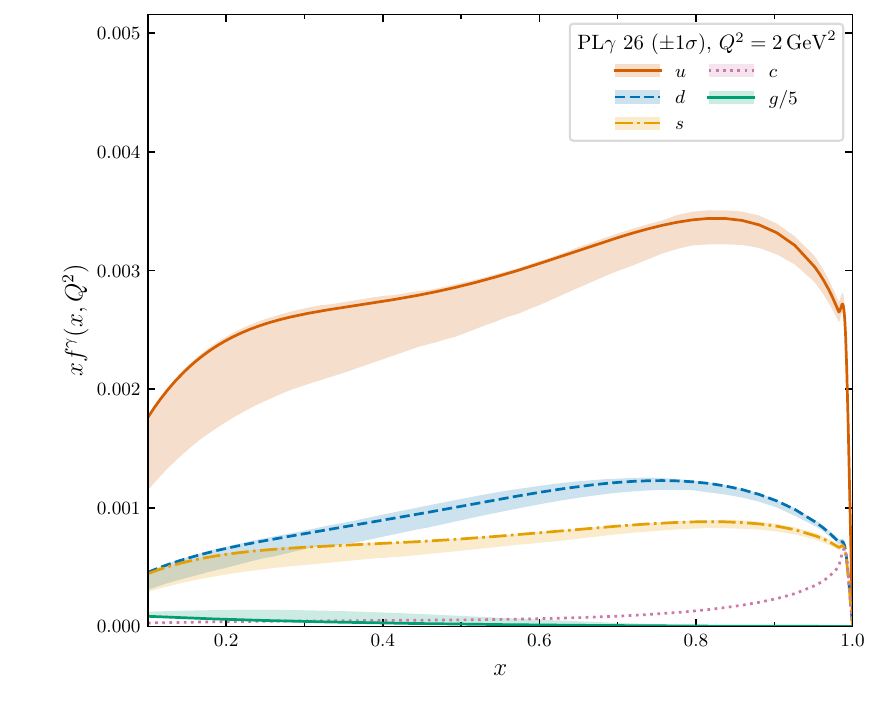}
        \label{fig:allflavs_q2_2}
    }
    \hfill
    \subfloat[\qsquared = \qty{5.0}{\GeV \squared}]{
        \includegraphics[width=0.48\columnwidth]{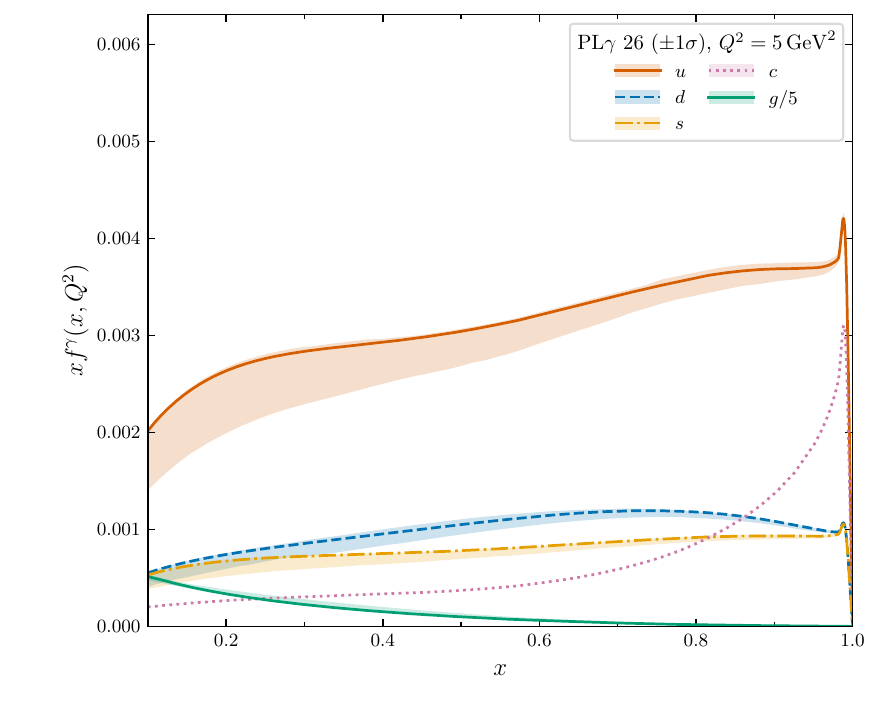}
        \label{fig:allflavs_q2_5}
    }

    \vspace{0.3cm}

    \subfloat[\qsquared = \qty{20.0}{\GeV \squared}]{
        \includegraphics[width=0.48\columnwidth]{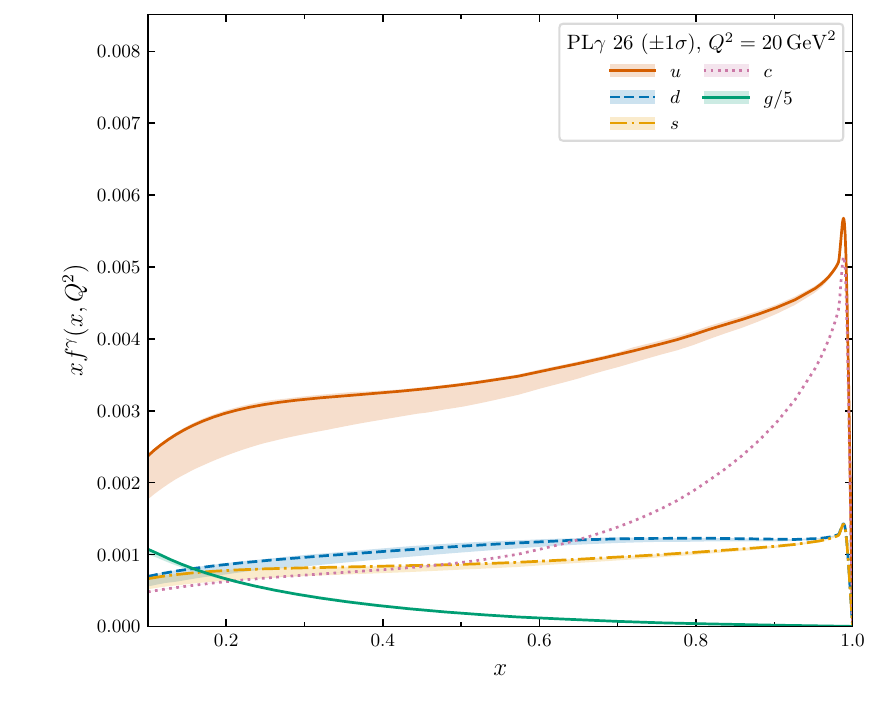}
        \label{fig:allflavs_q2_20}
    }
    \hfill
    \subfloat[\qsquared = \qty{50.0}{\GeV \squared}]{
        \includegraphics[width=0.48\columnwidth]{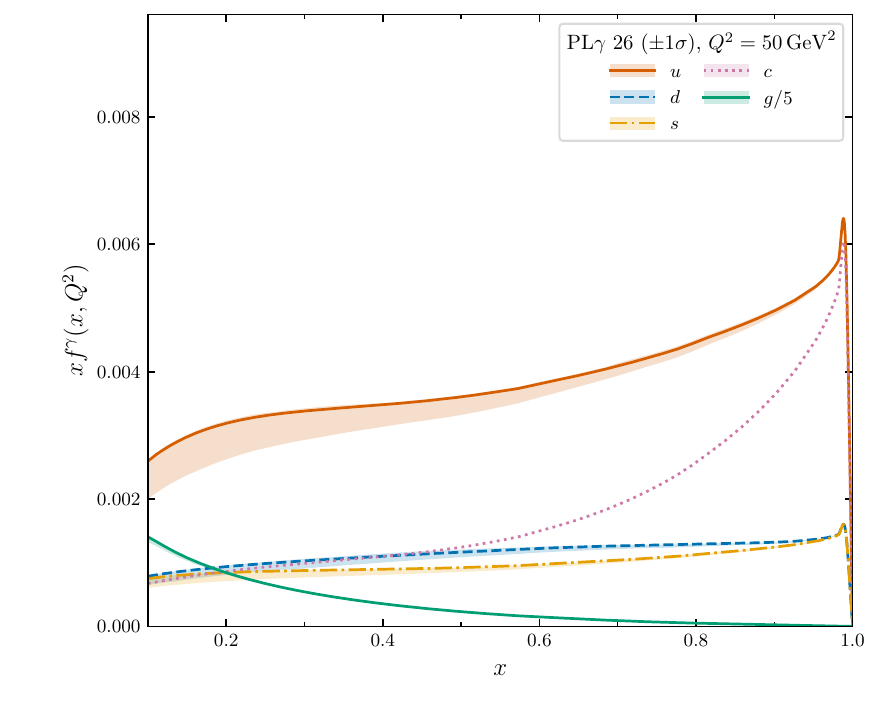}
        \label{fig:allflavs_q2_50}
    }

    \caption{All \plgamma flavours at different \qsquared values, with the central fit
      and the 68\,\% interval of the replicas. The
      $c$ PDF error band is too small to be visible, as charm-quark
      content arises predominantly via perturbative $\gamma$ splitting
      in the evolution, and is insensitive to the replica variations
      of parton content.}
    \label{fig:allflavs_q2_comp}
\end{figure}

Figure~\ref{fig:allflavs_q2_comp} shows the parton distributions of the fitted set, \plgamma, at
$\qsquared = 2$, $5$, $20$ and \qty{50}{\GeV\squared}.
The up quark is the largest quark distribution, as expected from its larger electric charge,
$e_u = \tfrac23$, while the down and strange quarks, with $e_{d,s} = -\tfrac13$, have similar distributions,
except at intermediate $x$, where the down quark is larger as its hadronic component peaks there.
The charm quark, generated radiatively above its threshold, grows with \qsquared, in particular at large $x$.
At $\qsquared \geq \qty{5}{\GeV\squared}$, the gluon dominates at small $x$ and falls steeply towards large $x$.

\begin{figure}[htbp]
    \centering
    \subfloat[Up quark]{
        \includegraphics[width=0.42\columnwidth]{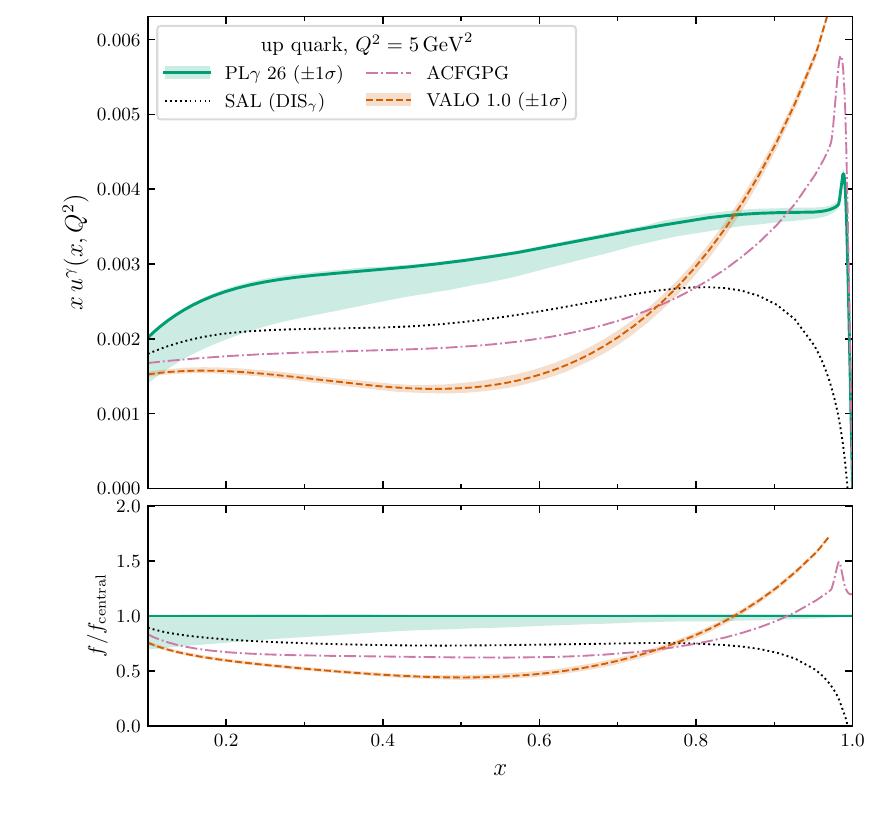}
    }
    \hfill
    \subfloat[Down quark]{
        \includegraphics[width=0.42\columnwidth]{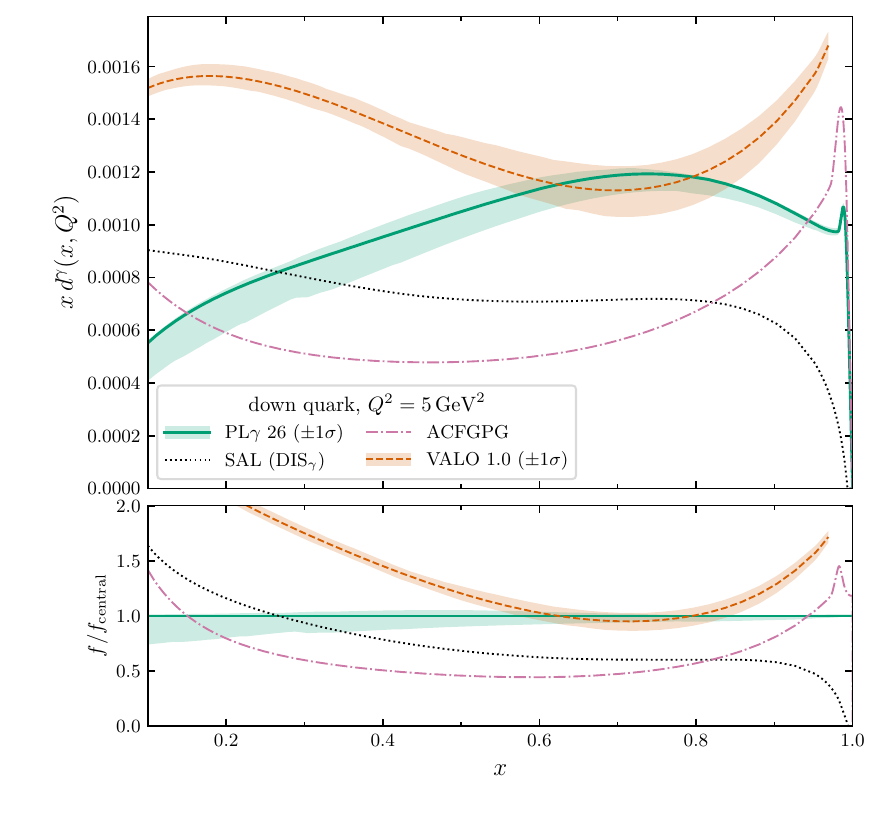}
    }

    \vspace{0.3cm}

    \subfloat[Strange quark]{
        \includegraphics[width=0.42\columnwidth]{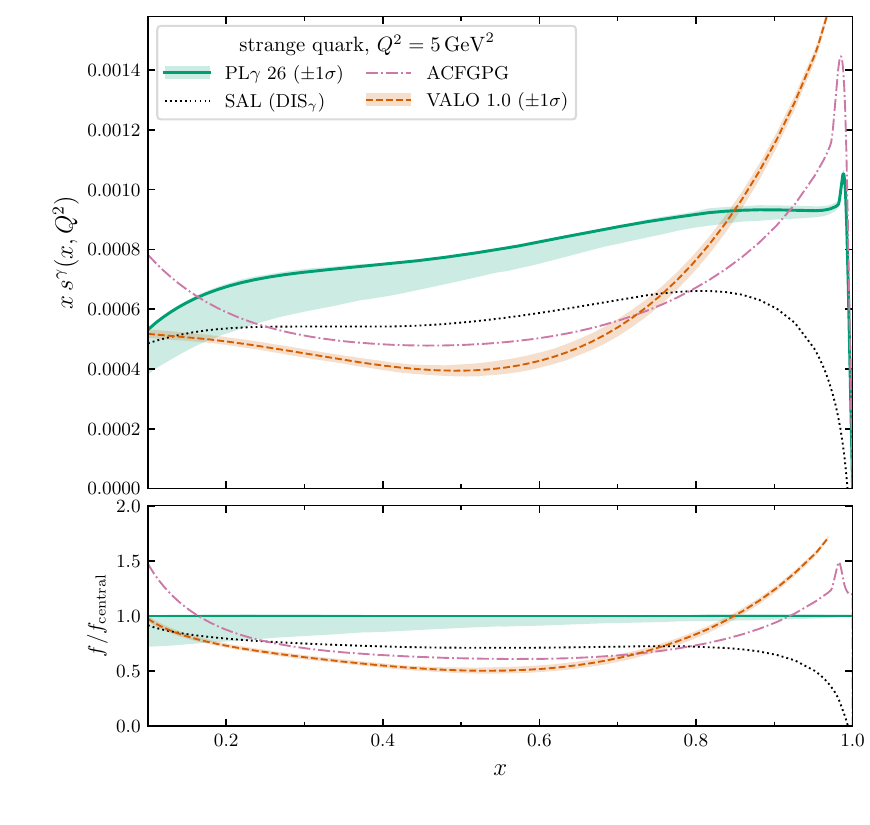}
    }
    \hfill
    \subfloat[Charm quark]{
        \includegraphics[width=0.42\columnwidth]{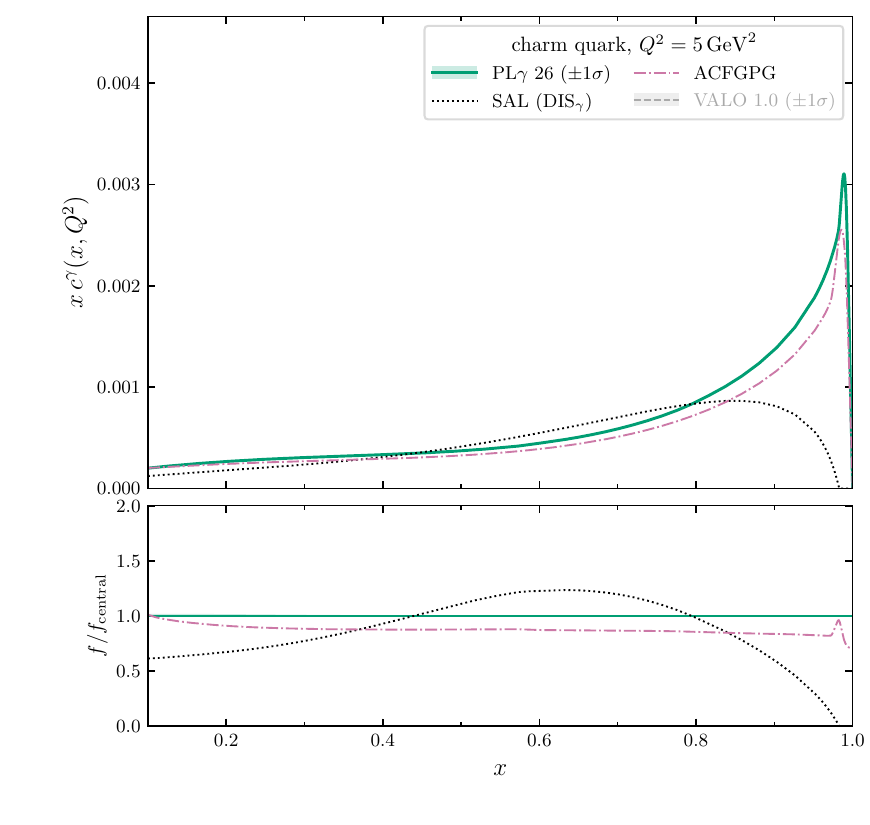}
    }

    \vspace{0.3cm}

    \subfloat[Gluon]{
        \includegraphics[width=0.42\columnwidth]{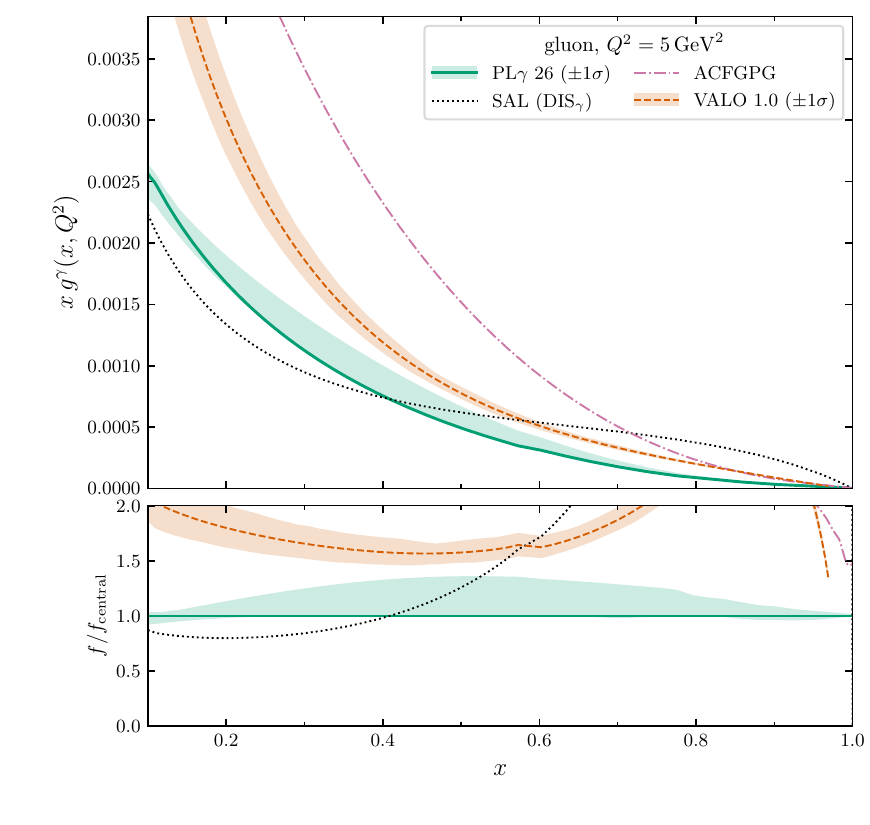}
    }
    \hfill
    \subfloat[Low-$x$ gluon]{
        \includegraphics[width=0.42\columnwidth]{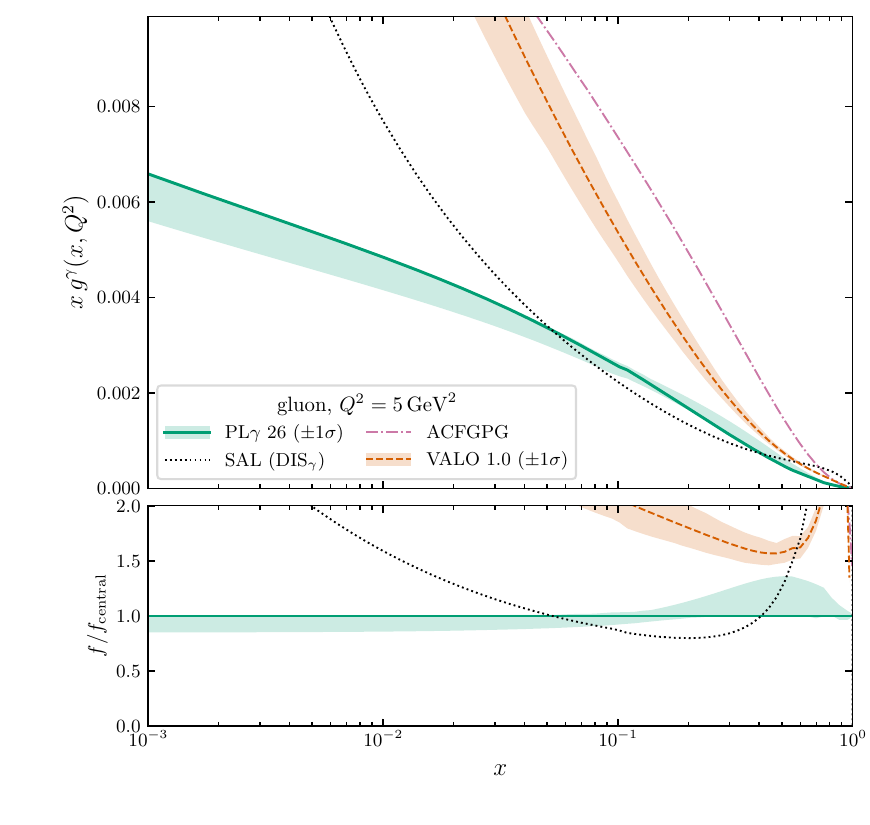}
    }
    \caption{\plgamma ($\pm1\sigma$, i.e.\ the 68\,\% interval of the replicas) compared with other photon-PDF sets at $\qsquared = \qty{5}{\GeV\squared}$
      for $x \geq 0.1$, and for $x \geq 10^{-3}$ in the low-$x$ gluon plot. The lower panels show the ratio to the central \plgamma.
      The charm-quark plot does not contain VALO~1.0 as the quark is negative at these \qsquared values due to
      near-threshold \DISgamma to \MSbar conversion.}
    \label{fig:pdf_comp_q2_5}
\end{figure}

In Fig.~\ref{fig:pdf_comp_q2_5}, we compare \plgamma with other photon-PDF sets
at $\qsquared = \qty{5}{\GeV\squared}$:
the NLO sets SAL~\cite{Slominski:2005bw} in the \DISgamma scheme, and ACFGP~\cite{Aurenche:1992sb}
and VALO\,1.0~\cite{Chithirasreemadam:2026mqp} in the \MSbar scheme.
GS96, shown in Fig.~\ref{fig:finalobs}, is not defined at this scale.

Towards $x = 1$, our sets diverge from the SAL PDFs. This is partly an effect of the factorisation scheme:
in the \DISgamma scheme, the quark distributions fall towards zero at large $x$, whereas in the
\MSbar scheme they continue to rise before falling steeply to zero.
Part of the instability in this region also comes from the interpolation of the \LHAPDF grids,
which we mitigate with a high density of grid points at large $x$.

At intermediate $x$, our up- and strange-quark PDFs lie 40--65\,\% above those of ACFGP and 60--120\,\% above
those of VALO\,1.0, and our down quark about a factor two above ACFGP, while it agrees with VALO\,1.0
for $x \approx 0.6$--$0.8$ and lies below it towards low $x$. Our gluon is smaller than in both \MSbar sets,
by a factor of 1.6--2 relative to VALO\,1.0 and about three relative to ACFGP.
At low $x$, our gluon diverges significantly from the other sets, as it rises much less steeply
towards small $x$ (Fig.~\ref{fig:pdf_comp_q2_5}). We note that our quark uncertainties
are also greater in the low-$x$ regime, growing from a few per cent at $x = 0.9$ to 25--30\,\% at $x = 0.1$,
simply because we do not have the data constraining this. The low-$x$ gluon band, of about 15\,\%, is narrower,
but reflects the restricted gluon parametrisation discussed below rather than a constraint from data.
To get a handle on this area, we must look towards the \ftwogamma measurements made at \LEP.
The bands are asymmetric, with the central quark PDFs at their upper edges and, above $x \approx 0.3$,
the central gluon at its lower edge, consistent with \apl, \ahad and \bhad of the central fit lying at the
upper edges of their sampling ranges and \ahadg at the lower edge (Table~\ref{tab:plg26-params}).
We note also that the charm PDF has small (invisible) error bands. As the charm is not parametrised at
\qzero, the central and replica bands rely on the DGLAP evolution itself. Most of the high-$x$ charm
is a result of the photon splitting $\gamma \rightarrow c\bar{c}$ and to a lesser extent the gluon splitting $g \rightarrow c\bar{c}$. Thus,
the replica variations in the charm that we would naturally expect to follow from variations in
the gluon and light quarks are not large enough to be visible on the plots, but this obviously does
not reflect a well-constrained charm quark.

The exponent \bhadg, which is primarily responsible for the low-$x$ tail, can, at very large negative values,
also affect the high-$x$ gluon region that we are sensitive to, but such values make the momentum
integral diverge, so we restricted the fit to $\bhad\!, \bhadg > -2.0$ (Table~\ref{tab:plg26-params}).
With this restriction, the central fit prefers \ahadg, the only other free gluon parameter, close to zero,
so that the gluon is generated almost entirely by the evolution and \bhadg is not constrained. We are content
with this because it allows our high-$x$ modelling to describe jet physics at \HERA and \LEP better than previous sets,
and demonstrates that our machinery does work, even with such parametrisation constraints. In future \plgamma work,
however, this does motivate a more complex parametrisation of the gluon (Section~\ref{sec:further}).

Differences between VALO\,1.0 and our \plgamma PDFs may be due to differences in the physics impact of
the observable selection.  While \ftwogamma, used in VALO\,1.0, fixes the quark content of the photon quite precisely,
it has practically no handle on its gluon content, which is therefore largely set by the parametrisation.
Conversely, using jet production only, as in the \plgamma set,
fixes the combined effect of gluons and quarks but it does not provide significant discrimination to separate
their relative impact.  As a consequence, the relatively larger gluon of VALO\,1.0, compared to \plgamma,
is compensated by a larger quark PDF in the latter.  This motivates further studies taking into
account both jet production and its more differential observables together with the more inclusive \ftwogamma,
as discussed in Section~\ref{sec:further}.

\section{Prospects for further development}
\label{sec:further}

As a new method, there are many possible avenues for further
investigation and refinement. Firstly, as noted in
Section~\ref{sec:param}, in this study we have used the
parametrisation of SAL~\cite{Slominski:2005bw}, in which the large-$x$
exponents are fixed to $\chad = 1$ and $\chadg = 3$. SAL took these
values from counting rules as their data did not reach $x \simeq 1$,
but the jet data used here have a degree of sensitivity to the
high-$x$ region.  An extended PDF tune could potentially achieve a
better fit by permitting the high-$x$ scaling exponents to vary,
freeing other elements of the parametrisation to achieve a better
description of lower-$x$ observables. We will pursue this in a
follow-up study, in which freeing \chadg will also test whether the
differences between our gluon and those of fits to \ftwogamma arise
from the gluon parametrisations, as discussed in
Section~\ref{sec:pdftune}. Similarly, the fixed
strange-suppression factor $r_s = 0.3$ would ideally also be
determined from data. The familiar issue of parametrisation bias also
raises the possibility of a more parametrisation-free basis,
e.g.~using the NNPDF approach of relatively unconstrained
non-parametric forms\footnote{Though encouraged towards classic PDF
structures by a preprocessing factor of the typical $x^\alpha
(1-x)^\beta$ character.}  for initial sampling~\cite{NNPDF:2014otw},
then rendered into a parametrised form for surrogate optimisation by
MC-to-Hessian conversion~\cite{Carrazza:2015aoa}.

A further step is a combined fit to the jet data and to the
\ftwogamma measurements~\cite{Nisius:1999cv} used in fits such as
VALO\,1.0~\cite{Chithirasreemadam:2026mqp}. These constrain the quark
densities directly, in particular at low $x$, while the jet data fix
the combined effect of quarks and gluons. Such a fit would therefore
separate the two, and show whether the compensation between quarks and
gluon discussed in Section~\ref{sec:pdftune} fully accounts for the
differences between \plgamma and fits to \ftwogamma, or whether there
is a tension between the two types of data.

Away from the PDF parametrisation, our approach taken to factorise the
MPI tuning from the PDF tune was for tractability: fit stability and
number of required sample points scale strongly with the number of
parameters. As the MPI tunes were performed using the SaS PDF, there is
an inconsistency between it and the final use-case. In practice,
MC generators are able to use separate PDFs for hard-process and MPI
modelling, avoiding the need to create an MPI tune for every
prospective resolved-photon PDF, but a single PDF suitable for all
aspects of MC generation would be both convenient and a test of model
completeness.

The MPI-tune factorisation between \HERA and \LEP could also be
refined, and combined with $pp$ and $p\bar{p}$ underlying-event and
minimum-bias data, to attempt a universal description of MPI able
to describe observables at $e^+e^-$, $ep$, and hadron colliders with a
single tune. While very interesting, this is a speculative goal and in
the current work the pragmatic approach was taken to find less
ambitious MPI configurations able to minimise biasing of the
resolved-photon PDF extraction.

Another prospect for future refinements in fits of this kind is to
incorporate estimations of MC-generator physics uncertainties into the
fit. At present, theory uncertainties are aggregated into the blanket
$\epsilon = 5\,\%$ regularisation term in Eq.~(\ref{eq:prof_chi2}).
But modern event-generator codes can provide far more detailed
breakdowns of sensitivity to modelling systematics such as
perturbative scale choices --- and increasingly also non-perturbative
modelling such as hadronisation and MPI~\cite{Pabst:2026pwv,Bierlich:2023fmh} ---
via event-weighting, a facility not yet used in this fit. Inclusion of
these systematic uncertainties would provide a further guard against
overfitting to a particular parton shower, or to less-constrained
elements of the non-PDF MC modelling, as well as in principle
permitting \emph{pairing} of resolved-photon PDF uncertainties with a
particular proton-PDF error set for $ep$ physics. This extension,
along with including an informed estimate of surrogate accuracy in the
$\chi^2$ definition, will require extensions in the \Professor
toolkit, but is aligned with planned developments for MC tuning.

\section{Conclusions}
\label{sec:concl}

In this work we have presented a new PDF fit for the resolved photon,
\plgamma, using measurements of jet production in photoproduction at
\HERA and $\gamma\gamma$ scattering at \LEP.

To arrive at the fit we take a novel approach. As these data typically
lie at rather low scales, non-factorisable corrections like
hadronisation or multi-parton interactions are not negligible. Hence,
instead of the usual approach of relying on post-processing of the
experimental data to ``subtract'' event-generator models of these
effects and produce data comparable with fixed-order partonic
calculations, we perform the PDF fit directly to the particle-level
data. This ensures forward application of the non-factorisable
modelling specific to each PDF configuration, and is made possible by
using the general-purpose event generator \Sherpa at NLO+PS accuracy.

As a first step, we tuned the \Sherpa multi-parton interactions model
separately to \HERA and \LEP jet data available in the \Rivet toolkit
(including new analyses added for this purpose), to avoid absorbing
such effects into the PDF. The resulting \HERA and \LEP MPI tunes are
themselves of interest for ongoing studies to quantify such effects
and optimise generator configurations for photoproduction and DIS
measurements, e.g.~in ultraperipheral heavy-ion collisions at the
LHC~\cite{Baltz:2007kq,Guzey:2018dlm,ATLAS:2024mvt}, at the upcoming
Electron--Ion Collider~\cite{Accardi:2012qut,AbdulKhalek:2021gbh,Meinzinger:2023xuf},
and in studies for FCC-ee~\cite{FCC:2018evy,FCC:2025lpp,Gehrmann:2025wbc}.

The resolved-photon PDF tune was then performed with the tuned \Sherpa
generator, on a sensitive subset of \HERA and \LEP data. Using the
6-parameter PDF ansatz from Ref.~\cite{Slominski:2005bw} at evolution
starting-scale $\qzero = \qty{1.3}{\GeV}$, points within the
6-dimensional parameter space were sampled
and evolved using \Apfelppp~\cite{Epping2023}, an appropriately
adapted version of the \Apfelpp library. This created a set of
\num{3000} random resolved-photon PDFs in the \LHAPDF data format;
treated as a single PDF set and using the NLO+PS reweighting mechanism
in \Sherpa, particle-level observable predictions were then
efficiently computed for all sampled PDF configurations within a
single MC-generator run. The \Professor MC-tuning framework was then
used first to construct surrogate models for each observable bin in
the 6-parameter PDF space, and secondly to optimise the fit to data
with \Professor's standard least-squares metric. A replica-fit
approach with 100~replicas from Gaussian-smeared data was used to
estimate the uncertainties.
The resulting functions are the first \MSbar resolved-photon PDFs fitted
to jet data from both \HERA and \LEP, and provide an error set for
quantitative uncertainty estimation. Compared with fits to \ftwogamma,
\plgamma has larger up- and strange-quark densities and a smaller gluon.
As \ftwogamma hardly constrains the gluon, while the jet data fix the
combined effect of quarks and gluons, we attribute this mainly to the
gluon parametrisations of those fits, compensated by larger quarks in
\plgamma; a combined fit to both types of data and a more flexible
parametrisation will test this reading.

This generator-based approach to PDF fitting is particularly valuable
in settings where data are sparse and the energy scales low enough to
be sensitive to non-factorisable effects. While the order of the fit
is restricted to NLO, this is competitive for the photon where NLO is
currently the standard, as well as by construction resolving any
debate about the formal compatibility of PDFs with MC
event-generation. These fits provide an important stepping stone to
updated phenomenological studies of jet production in ultraperipheral
collisions at the LHC, and in photoproduction at the EIC.

This fit uses complementary data compared to the fit performed in
Ref.~\cite{Chithirasreemadam:2026mqp}, which was performed to
$F_2^\gamma$ measurements. To our knowledge, it is the first
photon-PDF fit to a wide range of jet measurements from different
experiments, combining ten photoproduction and $\gamma\gamma$ analyses
from H1 and ZEUS at \HERA and from L3 and OPAL at \LEP. Earlier fits
were constrained mainly by \ftwogamma, adding at most a few jet
measurements from a single collider~\cite{Gordon:1996pm,Slominski:2005bw}. The resulting
set, \plgamma, and its replicas are provided in the \LHAPDF format as
\texttt{PLgamma26}, with member~0 the central fit to the unsmeared data and
members 1--100 the replicas. Until the MPI are tuned with \plgamma
itself (Section~\ref{sec:further}), we recommend using SaS\,1M with the tunes of
Table~\ref{tab:mpi_results} for the MPI, as in our fit.

This work, while not yet comparable to the scale and depth of global
PDF fits and with a number of immediate avenues identified for further
investigation, serves both to demonstrate the practical viability of
particle-level PDFs via the technical advances in modern MC event
generators and to highlight that a combined PDF- and generator-tuning
approach may be the most robust and practically useful way to define
PDFs for future low-scale physics at EIC and other facilities.

\section*{Acknowledgements}

\noindent Our thanks to Valerio Bertone for advice and assistance with the \Apfelpp
code, and obtaining a photon-PDF evolution with current tools. We also thank
Claire Gwenlan for advice and assistance with the ZEUS 3- and 4-jet analysis,
and Ilkka Helenius for providing \Pythia configurations and the preliminary \MSbar
VALO PDF.

\medskip

\noindent Funding acknowledgements for each author follow:
\begin{itemize}
\item AB acknowledges funding via the STFC experimental Consolidated
Grants programme (grant number ST/K001205/1), and was further supported
in this work by the STFC SWIFT-HEP programme, ST/Y005562/1.

\item FK acknowledges funding by STFC under grant agreement ST/P006744/1.

\item PM has been supported by the Swiss National Science Foundation (SNSF)
under contract 240015 and by the European Research Council (ERC)
under the European Union's Horizon 2020 research and innovation programme
grant agreement 101019620 (ERC Advanced Grant TOPUP).

\item VAN acknowledges a PhD studentship funded by the STFC,
grant number ST/Y509474/1, and the sub-department
of particle physics, University of Oxford.
\end{itemize}

\bibliography{photonpdf.bib}

\clearpage

\begin{appendix}
\numberwithin{equation}{section}

\section{Observable selection and weights for the MPI tunes}
\label{sec:AppMPI}

\bgroup
\centering\small
\begin{tabular}{llp{0.36\linewidth}rll}
\toprule
Histograms & Observable & Condition & Bins & Selected range & Weight \\
\midrule
\multicolumn{6}{l}{\texttt{H1\_2002\_I581409} (14 histograms, 67 bins selected)}\\
d01-x01-y01 & $M_{jj}$ &  & 6/6 & $[45, 180]$ & 1 \\
d02-x01-y01 & $\langle E_\mathrm{T}\rangle$ &  & 4/4 & $[20, 80]$ & 1 \\
d03-x01-y01 & $E_\mathrm{T}^{\max}$ &  & 4/4 & $[25, 80]$ & 1 \\
d04 (2) & $\bar\eta$ & \raggedright $25<E_\mathrm{T}<35\,\mathrm{GeV}$ / $35<E_\mathrm{T}<80\,\mathrm{GeV}$, $0.1<y<0.5$ & 9/9 & $[0.6, 2.5]$ & 1 \\
d05 (2) & $\bar\eta$ & \raggedright $25<E_\mathrm{T}<35\,\mathrm{GeV}$ / $35<E_\mathrm{T}<80\,\mathrm{GeV}$, $0.5<y<0.9$ & 10/10 & $[0, 2.5]$ & 1 \\
d06 (2) & \xgammaobs &  & 6/8 & $[0.1, 0.85]$ & 1 \\
d07-x01-y01 & $x_p^{\mathrm{obs}}$ & \raggedright low \xgammaobs & 5/5 & $[0.05, 0.6]$ & 1 \\
d07-x01-y02 & $x_p^{\mathrm{obs}}$ & \raggedright high \xgammaobs & 5/5 & $[0.05, 0.6]$ & 0 \\
d08 (2) & \xgammaobs & \raggedright low $E_\mathrm{T}$ / high $E_\mathrm{T}$ & 8/10 & $[0.1, 0.85]$ & 1 \\
d09-x01-y01 & $\vert\cos\theta^*\vert$ & \raggedright all $M_{jj}$, resolved & 7/7 & $[0, 0.7]$ & 1 \\
d09-x01-y02 & $\vert\cos\theta^*\vert$ & \raggedright all $M_{jj}$, direct & 7/7 & $[0, 0.7]$ & 0 \\
d10-x01-y01 & $\vert\cos\theta^*\vert$ & \raggedright $M_{jj}>65\,\mathrm{GeV}$, resolved & 8/8 & $[0, 0.85]$ & 1 \\
d10-x01-y02 & $\vert\cos\theta^*\vert$ & \raggedright $M_{jj}>65\,\mathrm{GeV}$, direct & 8/8 & $[0, 0.85]$ & 0 \\
\midrule
\multicolumn{6}{l}{\texttt{ZEUS\_2007\_I753991} (10 histograms, 62 bins selected)}\\
d01-x01-y01 & $E_\mathrm{T}$ & \raggedright direct & 8/8 & $[17.5, 90.5]$ & 0 \\
d02-x01-y01 & $E_\mathrm{T}$ & \raggedright resolved & 7/7 & $[17.5, 70.5]$ & 1 \\
d03-x01-y01 & $E_\mathrm{T}^{\mathrm{jet1}}$ & \raggedright direct & 7/7 & $[20, 90]$ & 0 \\
d04-x01-y01 & $E_\mathrm{T}^{\mathrm{jet1}}$ & \raggedright resolved & 6/6 & $[20, 70]$ & 1 \\
d05-x01-y01 & $\bar\eta$ & \raggedright direct & 6/6 & $[-0.5, 2.5]$ & 0 \\
d06-x01-y01 & $\bar\eta$ & \raggedright resolved & 6/6 & $[0, 2.75]$ & 1 \\
d07-x01-y01 & $x_p^{\mathrm{obs}}$ & \raggedright direct & 11/11 & $[0, 1]$ & 0 \\
d08-x01-y01 & $x_p^{\mathrm{obs}}$ & \raggedright resolved & 11/11 & $[0, 1]$ & 1 \\
d09-x01-y01 & $\Delta\phi$ & \raggedright direct & 5/5 & $[1.83, 3.14]$ & 0 \\
d10-x01-y01 & $\Delta\phi$ & \raggedright resolved & 7/7 & $[0, 3.14]$ & 1 \\
d11--d14 (4) & $x_p^{\mathrm{obs}}$ & \raggedright direct & 14/14 & $[0, 0.5]$ & 0 \\
d15--d18 (4) & $x_p^{\mathrm{obs}}$ & \raggedright resolved & 18/18 & $[0, 1]$ & 1 \\
d19-x01-y01 & \xgammaobs &  & 7/8 & $[0.1, 0.8]$ & 1 \\
\midrule
\multicolumn{6}{l}{\texttt{ZEUS\_1997\_I450085} (0 histograms, 0 bins selected)}\\
d01--d04 (4) & $\bar\eta$ & \raggedright $E_\mathrm{T}^{\mathrm{jet}}>6\,\mathrm{GeV}$ / $E_\mathrm{T}^{\mathrm{jet}}>8\,\mathrm{GeV}$ / $E_\mathrm{T}^{\mathrm{jet}}>11\,\mathrm{GeV}$ / $E_\mathrm{T}^{\mathrm{jet}}>15\,\mathrm{GeV}$, all \xgammaobs & 29/29 & $[-1.125, 1.625]$ & 0 \\
d21--d24 (4) & $\bar\eta$ & \raggedright $E_\mathrm{T}^{\mathrm{jet}}>6\,\mathrm{GeV}$ / $E_\mathrm{T}^{\mathrm{jet}}>8\,\mathrm{GeV}$ / $E_\mathrm{T}^{\mathrm{jet}}>11\,\mathrm{GeV}$ / $E_\mathrm{T}^{\mathrm{jet}}>15\,\mathrm{GeV}$, direct & 25/25 & $[-1.125, 1.625]$ & 0 \\
d25--d28 (4) & $\bar\eta$ & \raggedright $E_\mathrm{T}^{\mathrm{jet}}>6\,\mathrm{GeV}$ / $E_\mathrm{T}^{\mathrm{jet}}>8\,\mathrm{GeV}$ / $E_\mathrm{T}^{\mathrm{jet}}>11\,\mathrm{GeV}$ / $E_\mathrm{T}^{\mathrm{jet}}>15\,\mathrm{GeV}$, mid \xgammaobs & 22/22 & $[-0.375, 1.625]$ & 0 \\
\bottomrule
\end{tabular}
\captionof{table}{Observables used in the \HERA MPI tune (set 1). The analysis and
  histogram names follow the \Rivet convention. The observable column
  documents the differential observable, and the
  condition/selected-range columns note uses of only part of the
  histogram. Histograms with weight~0 are for monitoring only, and did
  not enter the fit.}  \egroup

\begin{table}[p]
\centering\small
\begin{tabular}{llp{0.36\linewidth}rll}
\toprule
Histograms & Observable & Condition & Bins & Selected range & Weight \\
\midrule
\multicolumn{6}{l}{\texttt{ZEUS\_2001\_I568665} (11 histograms, 56 bins selected)}\\
d01-x01-y01 & $\vert\cos\theta^*\vert$ & \raggedright resolved & 4/8 & $[0.4, 0.8]$ & 1 \\
d01-x01-y02 & $\vert\cos\theta^*\vert$ & \raggedright direct & 8/8 & $[0, 0.8]$ & 0 \\
d02--d07 (6) & $E_\mathrm{T}^{\mathrm{jet1}}$ & \raggedright direct, $\eta$ bin & 38/38 & $[14, 75]$ & 0 \\
d08--d13 (6) & $E_\mathrm{T}^{\mathrm{jet1}}$ & \raggedright resolved, $\eta$ bin & 32/32 & $[14, 65]$ & 1 \\
d14--d19 (6) & $\eta^{\mathrm{jet2}}$ & \raggedright direct, $\eta$ bin & 41/41 & $[-1, 2.4]$ & 0 \\
d20--d23 (4) & \xgammaobs & \raggedright $14<E_\mathrm{T}^{\mathrm{jet1}}<17\,\mathrm{GeV}$ / $17<E_\mathrm{T}^{\mathrm{jet1}}<25\,\mathrm{GeV}$ / $25<E_\mathrm{T}^{\mathrm{jet1}}<35\,\mathrm{GeV}$ / $35<E_\mathrm{T}^{\mathrm{jet1}}<90\,\mathrm{GeV}$ & 20/24 & $[0.1, 0.8]$ & 1 \\
\midrule
\multicolumn{6}{l}{\texttt{ZEUS\_2007\_I756660} (19 histograms, 192 bins selected)}\\
d01-x01-y01 & $M_{3j}$ &  & 11/11 & $[25, 160]$ & 1 \\
d02-x01-y01 & $M_{4j}$ &  & 11/11 & $[25, 160]$ & 0 \\
d03 (2) & \xgammaobs & \raggedright $25<M_{3j}<50\,\mathrm{GeV}$ / $M_{3j}>50\,\mathrm{GeV}$ & 13/17 & $[0, 0.85]$ & 1 \\
d04 (2) & \xgammaobs & \raggedright low $M_{4j}$ / high $M_{4j}$ & 17/17 & $[0, 1]$ & 0 \\
d05 (2) & $y$ & \raggedright low $M_{3j}$ / high $M_{3j}$ & 18/18 & $[0.2, 0.85]$ & 1 \\
d06 (2) & $y$ & \raggedright low $M_{4j}$ / high $M_{4j}$ & 18/18 & $[0.2, 0.85]$ & 0 \\
d07 (2) & $E_\mathrm{T}^{\mathrm{jet1}}$ & \raggedright low $M_{3j}$ / high $M_{3j}$ & 19/19 & $[6, 71]$ & 1 \\
d08 (2) & $E_\mathrm{T}^{\mathrm{jet2}}$ & \raggedright low $M_{3j}$ / high $M_{3j}$ & 16/16 & $[6, 60.8]$ & 1 \\
d09 (2) & $E_\mathrm{T}^{\mathrm{jet3}}$ & \raggedright low $M_{3j}$ / high $M_{3j}$ & 12/12 & $[6, 40]$ & 1 \\
d10 (2) & $E_\mathrm{T}^{\mathrm{jet1}}$ & \raggedright low $M_{4j}$ / high $M_{4j}$ & 16/16 & $[6, 60.8]$ & 0 \\
d11 (2) & $E_\mathrm{T}^{\mathrm{jet2}}$ & \raggedright low $M_{4j}$ / high $M_{4j}$ & 13/13 & $[6, 43.9]$ & 0 \\
d12 (2) & $E_\mathrm{T}^{\mathrm{jet3}}$ & \raggedright low $M_{4j}$ / high $M_{4j}$ & 9/9 & $[6, 26]$ & 0 \\
d13 (2) & $E_\mathrm{T}^{\mathrm{jet4}}$ & \raggedright low $M_{4j}$ / high $M_{4j}$ & 6/6 & $[6, 20]$ & 0 \\
d14 (2) & $\eta^{\mathrm{jet1}}$ & \raggedright low $M_{3j}$ / high $M_{3j}$ & 26/26 & $[-1.686, 2.4]$ & 1 \\
d15 (2) & $\eta^{\mathrm{jet2}}$ & \raggedright low $M_{3j}$ / high $M_{3j}$ & 27/27 & $[-2, 2.4]$ & 1 \\
d16 (2) & $\eta^{\mathrm{jet3}}$ & \raggedright low $M_{3j}$ / high $M_{3j}$ & 28/28 & $[-2, 2.4]$ & 1 \\
d17 (2) & $\eta^{\mathrm{jet1}}$ & \raggedright low $M_{4j}$ / high $M_{4j}$ & 24/24 & $[-1.371, 2.4]$ & 0 \\
d18 (2) & $\eta^{\mathrm{jet2}}$ & \raggedright low $M_{4j}$ / high $M_{4j}$ & 25/25 & $[-1.686, 2.4]$ & 0 \\
d19 (2) & $\eta^{\mathrm{jet3}}$ & \raggedright low $M_{4j}$ / high $M_{4j}$ & 25/25 & $[-1.686, 2.4]$ & 0 \\
d20 (2) & $\eta^{\mathrm{jet4}}$ & \raggedright low $M_{4j}$ / high $M_{4j}$ & 25/25 & $[-1.686, 2.4]$ & 0 \\
d21 (2) & $\cos\psi_3$ & \raggedright low $M_{3j}$ / high $M_{3j}$ & 22/22 & $[-1, 1]$ & 1 \\
\midrule
\multicolumn{6}{l}{\texttt{ZEUS\_2012\_I1116258} (10 histograms, 94 bins selected)}\\
d01-x01-y01 & $E_\mathrm{T}$ & \raggedright $\eta$ bin & 9/9 & $[17, 95]$ & 1 \\
d02--d03 (2) & $\eta$ & \raggedright $E_\mathrm{T}>17\,\mathrm{GeV}$ / $E_\mathrm{T}>21\,\mathrm{GeV}$ & 26/26 & $[-0.75, 2.5]$ & 1 \\
d04--d08 (5) & $E_\mathrm{T}$ & \raggedright $\eta$ bin & 41/41 & $[17, 95]$ & 1 \\
d09--d10 (2) & $E_\mathrm{T}$ & \raggedright anti-$k_t$ / SIScone & 18/18 & $[17, 95]$ & 1 \\
d11--d12 (2) & $\eta$ & \raggedright $E_\mathrm{T}>17\,\mathrm{GeV}$, anti-$k_t$ / SIScone & 26/26 & $[-0.75, 2.5]$ & 0 \\
\bottomrule
\end{tabular}
\caption{Observables used in the \HERA MPI tune (set 2). The analysis and
  histogram names follow the \Rivet convention. The observable column
  documents the differential observable, and the
  condition/selected-range columns note uses of only part of the
  histogram. Histograms with weight~0 are for monitoring only, and did
  not enter the fit.}
\end{table}

\begin{table}[p]
\centering\small
\begin{tabular}{llp{0.36\linewidth}rll}
\toprule
Histograms & Observable & Condition & Bins & Selected range & Weight \\
\midrule
\multicolumn{6}{l}{\texttt{H1\_2006\_I711847} (9 histograms, 43 bins selected)}\\
d01-x01-y01 & $\vert\cos\theta^*\vert$ & \raggedright resolved & 8/8 & $[0, 0.85]$ & 1 \\
d01-x01-y02 & $\vert\cos\theta^*\vert$ & \raggedright direct & 8/8 & $[0, 0.85]$ & 0 \\
d02-x01-y01 & $\vert\cos\theta^*\vert$ & \raggedright $M_{jj}$ cut, resolved & 8/8 & $[0, 0.85]$ & 1 \\
d02-x01-y02 & $\vert\cos\theta^*\vert$ & \raggedright $M_{jj}$ cut, direct & 8/8 & $[0, 0.85]$ & 0 \\
d03-x01-y01 & \xgammaobs &  & 4/4 & $[0.3, 1]$ & 0 \\
d03-x01-y02 & \xgammaobs &  & 4/5 & $[0.1, 0.85]$ & 1 \\
d04--d06 (3) & $x_p^{\mathrm{obs}}$ & \raggedright resolved & 8/13 & $[0.05, 0.32]$ & 1 \\
d07--d09 (3) & $x_p^{\mathrm{obs}}$ & \raggedright direct & 13/13 & $[0.05, 0.7]$ & 0 \\
d10--d12 (3) & $E_\mathrm{T}^{\max}$ & \raggedright resolved & 15/15 & $[25, 80]$ & 1 \\
d13--d15 (3) & $E_\mathrm{T}^{\max}$ & \raggedright direct & 16/16 & $[25, 80]$ & 0 \\
\bottomrule
\end{tabular}
\caption{Observables used in the \HERA MPI tune (set 3). The analysis and
  histogram names follow the \Rivet convention. The observable column
  documents the differential observable, and the
  condition/selected-range columns note uses of only part of the
  histogram. Histograms with weight~0 are for monitoring only, and did
  not enter the fit.}
\end{table}

\begin{table}[p]
\centering\small
\begin{tabular}{llp{0.36\linewidth}rll}
\toprule
Histograms & Observable & Condition & Bins & Selected range & Weight \\
\midrule
\multicolumn{6}{l}{\texttt{L3\_2004\_I661114} (1 histogram, 9 bins selected)}\\
d01-x01-y01 & $p_\mathrm{T}^{\mathrm{jet}}$ & \raggedright $\vert\eta\vert<1$ & 9/9 & $[3, 50]$ & 1 \\
\midrule
\multicolumn{6}{l}{\texttt{OPAL\_2008\_I754316} (2 histograms, 12 bins selected)}\\
d01--d02 (2) & $p_\mathrm{T}^{\mathrm{jet}}$ & \raggedright $\vert\eta\vert<1$ / $\vert\eta\vert<1.5$ & 12/12 & $[5, 40]$ & 1 \\
\midrule
\multicolumn{6}{l}{\texttt{OPAL\_2003\_I611415} (19 histograms, 113 bins selected)}\\
d01--d02 (2) & $\vert\cos\theta^*\vert$ & \raggedright $x_\gamma^\pm>0.75$ / $x_\gamma^\pm<0.75$ & 16/16 & $[0, 0.8]$ & 0 \\
d03--d05 (3) & $\bar{E}_\mathrm{T}$ & \raggedright all $x_\gamma^\pm$ / $x_\gamma^+$ or $x_\gamma^-<0.75$ / $x_\gamma^+$ and $x_\gamma^-<0.75$ & 15/15 & $[5, 25]$ & 1 \\
d06--d08 (3) & \xgammaobs & \raggedright $5<\bar{E}_\mathrm{T}<7\,\mathrm{GeV}$ / $7<\bar{E}_\mathrm{T}<11\,\mathrm{GeV}$ / $11<\bar{E}_\mathrm{T}<25\,\mathrm{GeV}$, all & 18/24 & $[0, 0.75]$ & 1 \\
d09--d10 (2) & \xgammaobs & \raggedright $5<\bar{E}_\mathrm{T}<7\,\mathrm{GeV}$ / $7<\bar{E}_\mathrm{T}<11\,\mathrm{GeV}$, $x_\gamma^+$ or $x_\gamma^-<0.75$ & 12/16 & $[0, 0.75]$ & 1 \\
d11--d12 (2) & \xgammaobs & \raggedright $5<\bar{E}_\mathrm{T}<7\,\mathrm{GeV}$ / $7<\bar{E}_\mathrm{T}<11\,\mathrm{GeV}$, $x_\gamma^+$ and $x_\gamma^-<0.75$ & 12/12 & $[0, 0.75]$ & 1 \\
d13--d15 (3) & $\ln\xgammaobs$ & \raggedright $5<\bar{E}_\mathrm{T}<7\,\mathrm{GeV}$, all / $x_\gamma^+$ or $x_\gamma^-<0.75$ / $x_\gamma^+$ and $x_\gamma^-<0.75$ & 12/15 & $[-1.65, -0.33]$ & 1 \\
d16--d17 (2) & $\vert\Delta\eta\vert$ & \raggedright $x_\gamma^+$ or $x_\gamma^-<0.75$ / $x_\gamma^+$ and $x_\gamma^-<0.75$ & 14/14 & $[0, 3.5]$ & 1 \\
d18--d19 (2) & $\min\vert\eta\vert$ & \raggedright $x_\gamma^+$ or $x_\gamma^-<0.75$ / $x_\gamma^+$ and $x_\gamma^-<0.75$ & 14/14 & $[0, 1.75]$ & 1 \\
d20--d21 (2) & $\max\vert\eta\vert$ & \raggedright $x_\gamma^+$ or $x_\gamma^-<0.75$ / $x_\gamma^+$ and $x_\gamma^-<0.75$ & 16/16 & $[0, 2]$ & 1 \\
\bottomrule
\end{tabular}
\caption{Observables used in the \LEP MPI tune. The analysis and histogram names follow the \Rivet convention. The observable column documents the differential observable, and the condition/selected-range columns note uses of only part of the histogram. Histograms with weight~0 are for monitoring only, and did not enter the fit.}
\end{table}

\begin{table}[htbp]
\centering
\begin{tabular}{lrrr}
\toprule
Analysis & Bins & Median $\sigma_\mathrm{MC}/\sigma_\mathrm{data}$ & Median $\sigma_\mathrm{MC}/\|\mathrm{MC}\|$ \\
\midrule
\multicolumn{4}{l}{\HERA}\\
\texttt{H1\_2002\_I581409} & 67 & 0.14 & 3.0\,\% \\
\texttt{H1\_2006\_I711847} & 43 & 0.32 & 4.3\,\% \\
\texttt{ZEUS\_2001\_I568665} & 56 & 0.90 & 8.8\,\% \\
\texttt{ZEUS\_2007\_I753991} & 62 & 0.40 & 4.9\,\% \\
\texttt{ZEUS\_2007\_I756660} & 192 & 0.85 & 11.9\,\% \\
\texttt{ZEUS\_2012\_I1116258} & 94 & 0.51 & 3.9\,\% \\
\midrule
\multicolumn{4}{l}{\LEP}\\
\texttt{L3\_2004\_I661114} & 5 & 0.22 & 2.1\,\% \\
\texttt{OPAL\_2003\_I611415} & 113 & 0.51 & 6.1\,\% \\
\texttt{OPAL\_2008\_I754316} & 11 & 0.15 & 1.7\,\% \\
\bottomrule
\end{tabular}
\caption{The statistical quality of the datasets used for the MPI tunes.}
\label{tab:mpi_mcstat}
\end{table}

\clearpage

\section{Observable selection and weights for the PDF tune}
\label{sec:AppPDF}
\bgroup
\centering\small
\begin{tabular}{llp{0.36\linewidth}rll}
\toprule
Histograms & Observable & Condition & Bins & Selected range & Weight \\
\midrule
\multicolumn{6}{l}{\texttt{H1\_2002\_I581409} (11 histograms, 49 bins selected)}\\
d01-x01-y01 & $M_{jj}$ &  & 4/6 & $[70, 180]$ & 1 \\
d02-x01-y01 & $\langle E_\mathrm{T}\rangle$ &  & 3/4 & $[30, 80]$ & 1 \\
d03-x01-y01 & $E_\mathrm{T}^{\max}$ &  & 4/4 & $[25, 80]$ & 1 \\
d04-x01-y01 & $\bar\eta$ & \raggedright $25<E_\mathrm{T}<35\,\mathrm{GeV}$, $0.1<y<0.5$ & 5/5 & $[0.6, 2.5]$ & 0 \\
d04-x01-y02 & $\bar\eta$ & \raggedright $35<E_\mathrm{T}<80\,\mathrm{GeV}$, $0.1<y<0.5$ & 4/4 & $[0.9, 2.5]$ & 1 \\
d05-x01-y01 & $\bar\eta$ & \raggedright $25<E_\mathrm{T}<35\,\mathrm{GeV}$, $0.5<y<0.9$ & 6/6 & $[0, 2.5]$ & 0 \\
d05-x01-y02 & $\bar\eta$ & \raggedright $35<E_\mathrm{T}<80\,\mathrm{GeV}$, $0.5<y<0.9$ & 4/4 & $[0.6, 2.1]$ & 1 \\
d06 (2) & \xgammaobs &  & 6/8 & $[0.1, 0.85]$ & 1 \\
d07-x01-y01 & $x_p^{\mathrm{obs}}$ & \raggedright low \xgammaobs & 5/5 & $[0.05, 0.6]$ & 1 \\
d07-x01-y02 & $x_p^{\mathrm{obs}}$ & \raggedright high \xgammaobs & 5/5 & $[0.05, 0.6]$ & 0 \\
d08-x01-y01 & \xgammaobs & \raggedright low $E_\mathrm{T}$ & 4/5 & $[0.1, 0.85]$ & 0 \\
d08-x01-y02 & \xgammaobs & \raggedright high $E_\mathrm{T}$ & 4/5 & $[0.1, 0.85]$ & 1 \\
d09-x01-y01 & $\vert\cos\theta^*\vert$ & \raggedright all $M_{jj}$, resolved & 7/7 & $[0, 0.7]$ & 1 \\
d09-x01-y02 & $\vert\cos\theta^*\vert$ & \raggedright all $M_{jj}$, direct & 7/7 & $[0, 0.7]$ & 0 \\
d10-x01-y01 & $\vert\cos\theta^*\vert$ & \raggedright $M_{jj}>65\,\mathrm{GeV}$, resolved & 8/8 & $[0, 0.85]$ & 1 \\
d10-x01-y02 & $\vert\cos\theta^*\vert$ & \raggedright $M_{jj}>65\,\mathrm{GeV}$, direct & 8/8 & $[0, 0.85]$ & 0 \\
\midrule
\multicolumn{6}{l}{\texttt{ZEUS\_2007\_I753991} (10 histograms, 59 bins selected)}\\
d01-x01-y01 & $E_\mathrm{T}$ & \raggedright direct & 8/8 & $[17.5, 90.5]$ & 0 \\
d02-x01-y01 & $E_\mathrm{T}$ & \raggedright resolved & 5/7 & $[27.5, 70.5]$ & 1 \\
d03-x01-y01 & $E_\mathrm{T}^{\mathrm{jet1}}$ & \raggedright direct & 7/7 & $[20, 90]$ & 0 \\
d04-x01-y01 & $E_\mathrm{T}^{\mathrm{jet1}}$ & \raggedright resolved & 5/6 & $[26, 70]$ & 1 \\
d05-x01-y01 & $\bar\eta$ & \raggedright direct & 6/6 & $[-0.5, 2.5]$ & 0 \\
d06-x01-y01 & $\bar\eta$ & \raggedright resolved & 6/6 & $[0, 2.75]$ & 1 \\
d07-x01-y01 & $x_p^{\mathrm{obs}}$ & \raggedright direct & 11/11 & $[0, 1]$ & 0 \\
d08-x01-y01 & $x_p^{\mathrm{obs}}$ & \raggedright resolved & 11/11 & $[0, 1]$ & 1 \\
d09-x01-y01 & $\Delta\phi$ & \raggedright direct & 5/5 & $[1.83, 3.14]$ & 0 \\
d10-x01-y01 & $\Delta\phi$ & \raggedright resolved & 7/7 & $[0, 3.14]$ & 1 \\
d11--d14 (4) & $x_p^{\mathrm{obs}}$ & \raggedright direct & 14/14 & $[0, 0.5]$ & 0 \\
d15--d18 (4) & $x_p^{\mathrm{obs}}$ & \raggedright resolved & 18/18 & $[0, 1]$ & 1 \\
d19-x01-y01 & \xgammaobs &  & 7/8 & $[0.1, 0.8]$ & 1 \\
\midrule
\multicolumn{6}{l}{\texttt{ZEUS\_1997\_I450085} (4 histograms, 14 bins selected)}\\
d01--d02 (2) & $\bar\eta$ & \raggedright $E_\mathrm{T}^{\mathrm{jet}}>6\,\mathrm{GeV}$ / $E_\mathrm{T}^{\mathrm{jet}}>8\,\mathrm{GeV}$, all \xgammaobs & 21/21 & $[-1.125, 1.625]$ & 0 \\
d03--d04 (2) & $\bar\eta$ & \raggedright $E_\mathrm{T}^{\mathrm{jet}}>11\,\mathrm{GeV}$ / $E_\mathrm{T}^{\mathrm{jet}}>15\,\mathrm{GeV}$, all \xgammaobs & 8/8 & $[-0.375, 1.625]$ & 1 \\
d21--d24 (4) & $\bar\eta$ & \raggedright $E_\mathrm{T}^{\mathrm{jet}}>6\,\mathrm{GeV}$ / $E_\mathrm{T}^{\mathrm{jet}}>8\,\mathrm{GeV}$ / $E_\mathrm{T}^{\mathrm{jet}}>11\,\mathrm{GeV}$ / $E_\mathrm{T}^{\mathrm{jet}}>15\,\mathrm{GeV}$, direct & 25/25 & $[-1.125, 1.625]$ & 0 \\
d25--d26 (2) & $\bar\eta$ & \raggedright $E_\mathrm{T}^{\mathrm{jet}}>6\,\mathrm{GeV}$ / $E_\mathrm{T}^{\mathrm{jet}}>8\,\mathrm{GeV}$, mid \xgammaobs & 16/16 & $[-0.375, 1.625]$ & 0 \\
d27--d28 (2) & $\bar\eta$ & \raggedright $E_\mathrm{T}^{\mathrm{jet}}>11\,\mathrm{GeV}$ / $E_\mathrm{T}^{\mathrm{jet}}>15\,\mathrm{GeV}$, mid \xgammaobs & 6/6 & $[0.125, 1.625]$ & 1 \\
\bottomrule
\end{tabular}
\captionof{table}{Observables used in the \plgamma tune. The analysis and histogram names follow the \Rivet convention.
The observable column documents the differential observable, and the condition/selected-range columns note uses of only part of the histogram.
 Histograms with weight~0 are for monitoring only, and did not enter the fit.}
\egroup

\begin{table}[p]
\centering\small
\begin{tabular}{llp{0.36\linewidth}rll}
\toprule
Histograms & Observable & Condition & Bins & Selected range & Weight \\
\midrule
\multicolumn{6}{l}{\texttt{ZEUS\_2001\_I568665} (9 histograms, 38 bins selected)}\\
d01-x01-y01 & $\vert\cos\theta^*\vert$ & \raggedright resolved & 4/8 & $[0.4, 0.8]$ & 1 \\
d01-x01-y02 & $\vert\cos\theta^*\vert$ & \raggedright direct & 8/8 & $[0, 0.8]$ & 0 \\
d02--d07 (6) & $E_\mathrm{T}^{\mathrm{jet1}}$ & \raggedright direct, $\eta$ bin & 38/38 & $[14, 75]$ & 0 \\
d08-x01-y01 & $E_\mathrm{T}^{\mathrm{jet1}}$ & \raggedright resolved, $\eta$ bin & 0/1 & none & 1 \\
d09--d13 (5) & $E_\mathrm{T}^{\mathrm{jet1}}$ & \raggedright resolved, $\eta$ bin & 21/31 & $[21, 65]$ & 1 \\
d14--d19 (6) & $\eta^{\mathrm{jet2}}$ & \raggedright direct, $\eta$ bin & 41/41 & $[-1, 2.4]$ & 0 \\
d20-x01-y01 & \xgammaobs & \raggedright $14<E_\mathrm{T}^{\mathrm{jet1}}<17\,\mathrm{GeV}$ & 7/8 & $[0.1, 0.8]$ & 0 \\
d21--d23 (3) & \xgammaobs & \raggedright $17<E_\mathrm{T}^{\mathrm{jet1}}<25\,\mathrm{GeV}$ / $25<E_\mathrm{T}^{\mathrm{jet1}}<35\,\mathrm{GeV}$ / $35<E_\mathrm{T}^{\mathrm{jet1}}<90\,\mathrm{GeV}$ & 13/16 & $[0.1, 0.8]$ & 1 \\
\midrule
\multicolumn{6}{l}{\texttt{ZEUS\_2007\_I756660} (40 histograms, 377 bins selected)}\\
d01-x01-y01 & $M_{3j}$ &  & 11/11 & $[25, 160]$ & 1 \\
d02-x01-y01 & $M_{4j}$ &  & 11/11 & $[25, 160]$ & 1$^\dagger$ \\
d03--d04 (4) & \xgammaobs & \raggedright $25<M_{3j}<50\,\mathrm{GeV}$ / $M_{3j}>50\,\mathrm{GeV}$ / low $M_{4j}$ / high $M_{4j}$ & 26/34 & $[0, 0.85]$ & 1$^\dagger$ \\
d05--d06 (4) & $y$ & \raggedright low $M_{3j}$ / high $M_{3j}$ / low $M_{4j}$ / high $M_{4j}$ & 36/36 & $[0.2, 0.85]$ & 1$^\dagger$ \\
d07 (2) & $E_\mathrm{T}^{\mathrm{jet1}}$ & \raggedright low $M_{3j}$ / high $M_{3j}$ & 19/19 & $[6, 71]$ & 1 \\
d08 (2) & $E_\mathrm{T}^{\mathrm{jet2}}$ & \raggedright low $M_{3j}$ / high $M_{3j}$ & 16/16 & $[6, 60.8]$ & 1 \\
d09 (2) & $E_\mathrm{T}^{\mathrm{jet3}}$ & \raggedright low $M_{3j}$ / high $M_{3j}$ & 12/12 & $[6, 40]$ & 1 \\
d10 (2) & $E_\mathrm{T}^{\mathrm{jet1}}$ & \raggedright low $M_{4j}$ / high $M_{4j}$ & 16/16 & $[6, 60.8]$ & 1$^\dagger$ \\
d11 (2) & $E_\mathrm{T}^{\mathrm{jet2}}$ & \raggedright low $M_{4j}$ / high $M_{4j}$ & 13/13 & $[6, 43.9]$ & 1$^\dagger$ \\
d12 (2) & $E_\mathrm{T}^{\mathrm{jet3}}$ & \raggedright low $M_{4j}$ / high $M_{4j}$ & 9/9 & $[6, 26]$ & 1$^\dagger$ \\
d13 (2) & $E_\mathrm{T}^{\mathrm{jet4}}$ & \raggedright low $M_{4j}$ / high $M_{4j}$ & 6/6 & $[6, 20]$ & 1$^\dagger$ \\
d14 (2) & $\eta^{\mathrm{jet1}}$ & \raggedright low $M_{3j}$ / high $M_{3j}$ & 26/26 & $[-1.686, 2.4]$ & 1 \\
d15 (2) & $\eta^{\mathrm{jet2}}$ & \raggedright low $M_{3j}$ / high $M_{3j}$ & 27/27 & $[-2, 2.4]$ & 1 \\
d16 (2) & $\eta^{\mathrm{jet3}}$ & \raggedright low $M_{3j}$ / high $M_{3j}$ & 28/28 & $[-2, 2.4]$ & 1 \\
d17 (2) & $\eta^{\mathrm{jet1}}$ & \raggedright low $M_{4j}$ / high $M_{4j}$ & 24/24 & $[-1.371, 2.4]$ & 1$^\dagger$ \\
d18 (2) & $\eta^{\mathrm{jet2}}$ & \raggedright low $M_{4j}$ / high $M_{4j}$ & 25/25 & $[-1.686, 2.4]$ & 1$^\dagger$ \\
d19 (2) & $\eta^{\mathrm{jet3}}$ & \raggedright low $M_{4j}$ / high $M_{4j}$ & 25/25 & $[-1.686, 2.4]$ & 1$^\dagger$ \\
d20 (2) & $\eta^{\mathrm{jet4}}$ & \raggedright low $M_{4j}$ / high $M_{4j}$ & 25/25 & $[-1.686, 2.4]$ & 1$^\dagger$ \\
d21 (2) & $\cos\psi_3$ & \raggedright low $M_{3j}$ / high $M_{3j}$ & 22/22 & $[-1, 1]$ & 1 \\
\midrule
\multicolumn{6}{l}{\texttt{ZEUS\_2012\_I1116258} (9 histograms, 57 bins selected)}\\
d01-x01-y01 & $E_\mathrm{T}$ & \raggedright $\eta$ bin & 6/9 & $[29, 95]$ & 1 \\
d02-x01-y01 & $\eta$ & \raggedright $E_\mathrm{T}>17\,\mathrm{GeV}$ & 13/13 & $[-0.75, 2.5]$ & 0 \\
d03-x01-y01 & $\eta$ & \raggedright $E_\mathrm{T}>21\,\mathrm{GeV}$ & 13/13 & $[-0.75, 2.5]$ & 1 \\
d04--d08 (5) & $E_\mathrm{T}$ & \raggedright $\eta$ bin & 26/41 & $[29, 95]$ & 1 \\
d09--d10 (2) & $E_\mathrm{T}$ & \raggedright anti-$k_t$ / SIScone & 12/18 & $[29, 95]$ & 1 \\
d11--d12 (2) & $\eta$ & \raggedright $E_\mathrm{T}>17\,\mathrm{GeV}$, anti-$k_t$ / SIScone & 26/26 & $[-0.75, 2.5]$ & 0 \\
\bottomrule
\end{tabular}
\caption{Observables used in the \plgamma tune. The analysis and histogram names follow the \Rivet convention.
The observable column documents the differential observable, and the condition/selected-range columns note uses of only part of the histogram.
 Histograms with weight~0 are for monitoring only, and did not enter the fit.
 Weights with 1$^\dagger$\ denote the use of \Professor's \texttt{normA} mode, where the fit is performed against the
 histograms, normalised to unit area --- that is, we perform a shape fit, ignoring the normalisation.}
\end{table}

\end{appendix}

\end{document}